\documentclass[sigconf, 10pt]{acmart}

\renewcommand\footnotetextcopyrightpermission[1]{} 
\usepackage{booktabs} 
\usepackage{pifont}		
\usepackage{graphicx}
\usepackage{epstopdf}
\usepackage{subcaption}
\usepackage[linesnumbered,algoruled,boxed,lined]{algorithm2e}

\setcopyright{rightsretained}

\acmConference[MOBICOM 2018]{}{}{}

\begin{document}
\title{Energy-Efficient Mobile Network I/O Optimization \\at the Application Layer}

\author{Kemal Guner\textsuperscript{*}, MD S Q Zulkar Nine\textsuperscript{*}, Tevfik Kosar\textsuperscript{*}, M. Fatih Bulut\textsuperscript{\ddag}}
\affiliation{%
  \institution{\textsuperscript{*\ }University at Buffalo (SUNY) \\ \textsuperscript{\ddag }IBM Thomas J. Watson Research Center \\ \{kemalgne, mdsqzulk, tkosar\}@buffalo.edu, mfbulut@us.ibm.com\\}  
}

\begin{abstract}
Mobile data traffic (cellular + WiFi) will exceed PC Internet traffic by 2020. As the number of smartphone users and the amount of data transferred per smartphone grow exponentially, limited battery power is becoming an increasingly critical problem for mobile devices which depend on the network I/O. Despite the growing body of research in power management techniques for the mobile devices at the hardware layer as well as the lower layers of the networking stack, there has been little work focusing on saving energy at the application layer for the mobile systems during network I/O. 
In this paper, to the best of our knowledge, we are first to provide an in-depth analysis of the effects of application-layer data transfer protocol parameters on the energy consumption of mobile phones. We propose a novel model, called FastHLA, that can achieve significant energy savings at the application layer during mobile network I/O without sacrificing the performance. In many cases, our model achieves performance increase and energy saving simultaneously. 
\end{abstract}


\maketitle

\section{Introduction}
\label{sec:intro}

It is estimated that smartphone mobile data traffic (cellular + WiFi) will reach 370 Exabytes per year by 2020, exceeding PC Internet traffic the first time in the history~\cite{Cisco_2016}. 
An average smartphone consumes between 300 -- 1200 milliwatts power~\cite{carroll2010analysis} depending on the type of applications it is running, and most of the energy in smartphone applications is spent for networked I/O. During an active data transfer, the cellular (i.e., GSM) and WiFi components of a smartphone consume more power than its CPU, RAM, and even LCD+graphics card at the highest brightness level \cite{pathak2012energy, carroll2010analysis}. Although the mobile data traffic and the amount of energy spent for it increase at a very fast pace, the battery capacities of smartphones do not increase at the same rate.

Limited battery power is becoming an increasingly critical problem for smartphones and mobile computing, and many techniques have been proposed in the literature to overcome this at different layers. 
At the physical layer, techniques were proposed to choose appropriate modulation, coding, and transmission power control schemes to improve energy efficiency of the mobile device~\cite{cianca2001improving, schurgers2001modulation, cui2004energy, singh1998pamas, takai2001effects}.
At the media access control (MAC) layer, several new energy-efficient MAC protocol designs were proposed~\cite{woesner1998power, ye2002energy, bharghavan1994macaw, woo2001transmission, krashinsky2005minimizing, krashinsky2005minimizing, nasipuri2000mac}.
At the network layer, low-power and scalable routing algorithms were developed~\cite{xu2001geography, chang2000energy, seada2004energy, singh1998power, toh2001maximum}.
At the transport layer, traffic shaping techniques~\cite{akella2001protocols} and new transport protocols~\cite{zorzi1999tcp, kravets2000application, akella2001protocols, haas1997mobile, chandra2002application} were proposed to exploit application-specific information and reduce power utilization.  

Despite the growing body of research in power management techniques for the lower layers of the mobile networking stack, there has been little work focusing on saving network I/O (data transfer) energy at the application layer. 
The most notable work in this area are: tuning the client playback buffer size during media streaming in order to minimize the total energy spent~\cite{bertozzi2002power}; using lossless compression techniques to minimize the amount of data transferred as well as the energy consumed on wireless devices ~\cite{xu2003impact}; and joint optimization of the application layer, data link layer, and physical layer of the protocol stack using an application-oriented objective function in order to improve multimedia quality and power consumption at the same time~\cite{khan2006application}.
We claim that significant amount of network I/O energy savings can be obtained at the application layer with no or minimal performance penalty. Although lower-layer network stack approaches are an important part of the solution, application-layer power management is another key to optimizing network I/O energy efficiency in mobile computing, as a complementary approach to the optimizations at the lower layers of the networking stack.

In this paper, we analyze the effects of different application-layer data transfer protocol parameters (such as the number of parallel data streams per file, the level of concurrent file transfers to fill the mobile network pipes, and the I/O request size) on mobile data transfer throughput and energy consumption. Then, we propose a novel model that can achieve significant energy savings at the application layer during mobile network I/O without sacrificing the performance. 

In summary, our contributions within this paper are the following: 

\begin{itemize}

\item To the best of our knowledge, we are first to provide an in depth analysis of the effects of application-layer data transfer protocol parameters on the energy consumption of mobile phones.

\item We propose a novel historical-data analysis based model, called FastHLA, that can achieve significant energy savings at the application layer during mobile network I/O without sacrificing the performance. 

\item We show that our FastHLA model can achieve significant energy savings  using only application-layer solutions at the mobile systems during data transfer with no or minimal performance penalty.

\item We also show that, in many cases, our FastHLA model can increase the performance and save energy simultaneously.

\end{itemize}

The rest of this paper is organized as follows: Section II presents background information on energy-aware tuning of application-layer data transfer protocol parameters and discusses the related work in this area; Section III provides the methodology of our analysis; Section IV presents an in-depth experimental analysis of the application-layer parameter effects on mobile data transfer performance and energy consumption; Section V introduces our novel FastHLA model and compares it to the competing approaches; and Section VI concludes the paper.

\section{Background}
\label{sec:background}

The majority of work on mobile device energy savings mostly focuses putting the devices to sleep during idle times~\cite{krashinsky2005minimizing, vallina2011erdos, schulman2010bartendr, vallina2013energy}. A recent study by Dogar et al.~\cite{Dogar2010} takes this approach to another step, and puts the device into sleep even during data transfer by exploiting the high-bandwidth wireless interface. They combine small gaps between packets into meaningful sleep intervals, thereby allowing the NIC as well as the device to doze off. Another track of study in this area focuses on switching among multiple radio interfaces in an attempt to reduce the overall power consumption of the mobile device~\cite{pering2006coolspots, correia2010challenges, balasubramanian2009energy, nika2015energy}.
These techniques are orthogonal to our application-layer protocol tuning approach and could be used together to achieve higher energy efficiency in the mobile systems.

The closest work to ours in the literature is the work by Bertozzi et al.~\cite{bertozzi2003transport}, in which they investigate the energy trade-off in mobile networking as a function of the TCP receive buffer size and show that the TCP buffering mechanisms can be exploited to significantly increase energy efficiency of the transport layer with minimum performance overheads.

In this work, we focus on the tuning of three different protocol parameters:  concurrency (the level of concurrent file transfers to fill the mobile network pipes), parallelism (the number of parallel data streams per file), and I/O request size.

{\em Concurrency} refers to sending multiple files simultaneously through the network using different data channels at the same time.
Most studies in this area do not take the data size and the network characteristics into consideration when setting the concurrency level~\cite{kosar04, Thesis_2005, Kosar09, JGrid_2012}. Liu et al.~\cite{R_Liu10} adapt the concurrency level based on the changes in the network traffic, but do not take into account other bottlenecks that can occur on the end systems. 

{\em Parallelism} sends different chunks of the same file using different data channels (i.e., TCP streams) at the same time 
and achieves high throughput by mimicking the behavior of individual streams and getting a higher share of the available bandwidth~\cite{R_Sivakumar00, R_Lee01, R_Balak98, R_Hacker05, R_Eggert00, R_Karrer06, R_Lu05, DADC_2008, DADC_2009, NDM_2011}. On the other hand, using too many simultaneous connections congests the network and the throughput starts dropping down. 
Predicting the optimal parallel stream number for a specific setting is a very challenging problem due to the dynamic nature of the interfering background traffic. 
Hacker et al. claimed that the total number of streams behaves like one giant stream that transfers in total capacity of each streams' achievable throughput~\cite{R_Hacker02}. However, this model only works for uncongested networks, since it accepts that packet loss ratio is stable and does not increase as the number of streams increases. 
Dinda et al. modeled the bandwidth of multiple streams as a partial second order polynomial which needs two different real-time throughput measurements to provide accurate predictions~\cite{R_Dinda05}. 

{\em I/O request size} is the size of request that application uses to perform I/O operation on storage device.
The I/O request size may have a big impact on the storage performance, and also on the end-to-end data transfer performance if the end system storage throughput is the main bottleneck. 

When used wisely, these parameters have a potential to improve the end-to-end data transfer performance at a great extent, but improper use of these parameters can also hurt the performance of the data transfers due to increased load at the end-systems and congested links in the network~\cite{NDM_2012, Grid_2008}. For this reason, it is crucial to find the best combination for these parameters with the least intrusion and overhead to the system resource utilization and power consumption.

\setlength\belowcaptionskip{5ex}
\begin{table*}[t]
\small
	\begin{centering}
		\begin{tabular}{ |r|r|r|r|r| }
			\hline
			\rule{0pt}{2.3ex}
			Producer & Google & Samsung &  Samsung  & Samsung \\
			\hline
			\rule{0pt}{2.3ex}
			Model & Nexus S & Galaxy Nexus N3 (L700) & Galaxy S4 & Galaxy S5  \\
			
			OS & Android 4.1.1 (API 16) & Android 4.3 (API 18) & Android 5.0.1 (API 21) & Android 5.0.1 (API 21)	\\
			
			CPU & 1.0 GHz Cortex-A8 & Dual-core 1.2 GHz & Quad-core 1.9 GHz Krait 300 & Quad-core 2.5 GHz Krait 400	\\
			
			Wifi & 802.11 b/g/n & 802.11 a/b/g/n & 802.11 a/b/g/n/ac & 802.11 a/b/g/n/ac	\\
			
			Storage & 16 GB	& 32 GB & 16 GB & 16 GB	\\
			
			Memory & 512 MB & 1 GB & 2 GB & 2 GB	\\
			\hline
			
		\end{tabular}
		\vspace{-5mm}
		\caption{Specifications of the mobile devices used in the experiments.} \label{tab:phonespecs}
	\end{centering}
	\vspace{-2mm}
\end{table*}

In the literature, several highly-accurate predictive models ~\cite{R_Yin11, R_Yildirim11, DISCS12, Cluster_2015} were developed which would require as few as three sampling points to provide very accurate predictions for the parallel stream number giving the highest transfer throughput for the wired networks. 
Yildirim et al. analyzed the combined effect of parallelism and concurrency on end-to-end data transfer throughput~\cite{TCC2015}.
Managed File Transfer (MFT) systems were proposed which used a subset of these parameters in an effort to improve the end-to-end data transfer throughput~\cite{WORLDS_2004, ScienceCloud_2013, globusonline, Royal_2011, IGI_2012}.
Alan et al. analyzed the effects of parallelism and concurrency on end-to-end data transfer throughput versus total energy consumption in wide-area wired networks in the context of GridFTP data transfers~\cite{Alan2015, Kosar_jrnl14}. {\em None of the existing work in this area studied the effects of these three parameters on the mobile energy consumption and the performance versus energy trade-offs of tuning these parameters in this context.}

\section{Methodology}
\label{sec:methodology}

In our analysis, we used a single-phase portable Yokogawa WT210 power meter, which provides highly accurate and fine granular power values (can measure DC and AC signals from 0.5 Hz to 100 kHz with an accuracy of 99.8\%) and is one of the accepted devices by the Standard Performance Evaluation Corporation (SPEC) power committee for power measurement and analysis purposes in the field~\cite{specOverview}. This power meter is used to measure the power consumption rates during the data transfers at the mobile client device. 

Prior to initiating any data transfer, we examined the base power state of each tested mobile device. 
To measure the base power state, we established a setting when the mobile device is in the ``on'' state with the screen is also on (always at the same brightness level), any communication interface other than the one being tested (i.e., Wifi or 4G LTE) is disabled, and a minimum number of necessary applications are running in the background. This setup ensured that the base power of the tested mobile device is both low and in a balanced state throughput the experiments. 

We designed a real time test environment with four different mobile devices (as specifications presented in Table~\ref{tab:phonespecs}). We tested both WiFi and 4G LTE connections in progress of data transfers on end-systems. 
To reduce the effect of number of active users and the effect of peak/off-peak hours during the transfer of datasets, we adopted a strategy of using different time frames for each of the same experiment settings, and take the average throughput and energy consumption values. We conducted all experiments at the same location and with the same distance and interference for objective analysis of the end-system devices. 

We run initial tests for all four mobile devices at different times of the day to obtain robust base power for each. With the help of these values, the total energy consumption during data transfers is calculated as follows:

\vspace{-3mm}

\begin{equation}
\hspace{-2.8cm}	
E_{t}  = E_{b} + E_{d}
\end{equation}
\vspace{-2mm}
\begin{equation}
E_{d}  = \int_{t_{start}}^{t_{end}} (P_{max}(t) - P_{b}(t)) \cdot dt
\label{eq:power}
\end{equation}

\noindent where,
\vspace{1mm}

\indent $\bullet$ $E_{t}$: Total energy consumption of data transfer \\
\indent $\bullet$ $E_{d}$: Dynamic energy consumption of data transfer \\
\indent $\bullet$ $E_{b}$: Base energy consumption of data transfer \\
\indent $\bullet$ $P_{max}$: Total power consumption  \\
\indent $\bullet$ $P_{b}$: Base power consumption before initiating the test \\
\indent $\bullet$ $t_{start}$: Data transfer start time \\
\indent $\bullet$ $t_{end}$: Data transfer end time \\

\setlength\belowcaptionskip{5ex}
\begin{table*}[t]
\small
	\begin{centering}
		\begin{tabular}{ r@{\hskip 0.5cm}r@{\hskip 0.5cm}r}
			\hline
			{\bf Dataset Name} &  {\bf Ave. File Size}  & {\bf Min-Max } \\
			\hline
			\rule{0pt}{2.3ex}
			HTML	&  112 KB 	& 56 KB - 155 KB 	\\
			IMAGE	&  2.7 MB 	& 2 MB - 3.2 MB 	\\		
			VIDEO-small	&  152 MB 	& 140 MB - 167 MB 	\\
			VIDEO-medium		&  3 GB		& 2.86 GB - 3.1 GB 		\\
			VIDEO-large	&  10GB 	& 9.7 GB - 10.2 GB 	\\
			\hline
		\end{tabular}
        \vspace{-2mm}
		\caption{Characteristics of the dataset used in the experiments.} \label{tab:dataset}
	\end{centering}
	\vspace{-2mm}
\end{table*}

Dynamic energy consumption $E_{d}$ in Equation \ref{eq:power} is established by taking integral of  subtract values of base power of device from total instantaneous power measured by power meter per second. All the energy consumption results presented in the paper refer to dynamic energy consumption as stated in Equation \ref{eq:power}. Since we aim to analyze the effect of application-layer parameters on energy consumption, we ignored the energy consumed when the device is idle.

We chose HTTP (Hypertext Transport Protocol) as the application-layer transfer protocol to test the impact of the parameters of interest on the end-to-end data transfer throughput as well as the energy consumption of the mobile client. The main reason for this choice is that HTTP is the de-facto transport protocol for Web services ranging from file sharing to media streaming, and the studies analyzing the Internet traffic~\cite{kellerman2016daily, Mirkovic2015, Czyz2014} show that HTTP accounts for 75\% of global mobile Internet traffic.

\begin{figure}[b]
	\captionsetup{justification=centering}
	\begin{center}
		\includegraphics[keepaspectratio=true,angle=0, width=85mm]{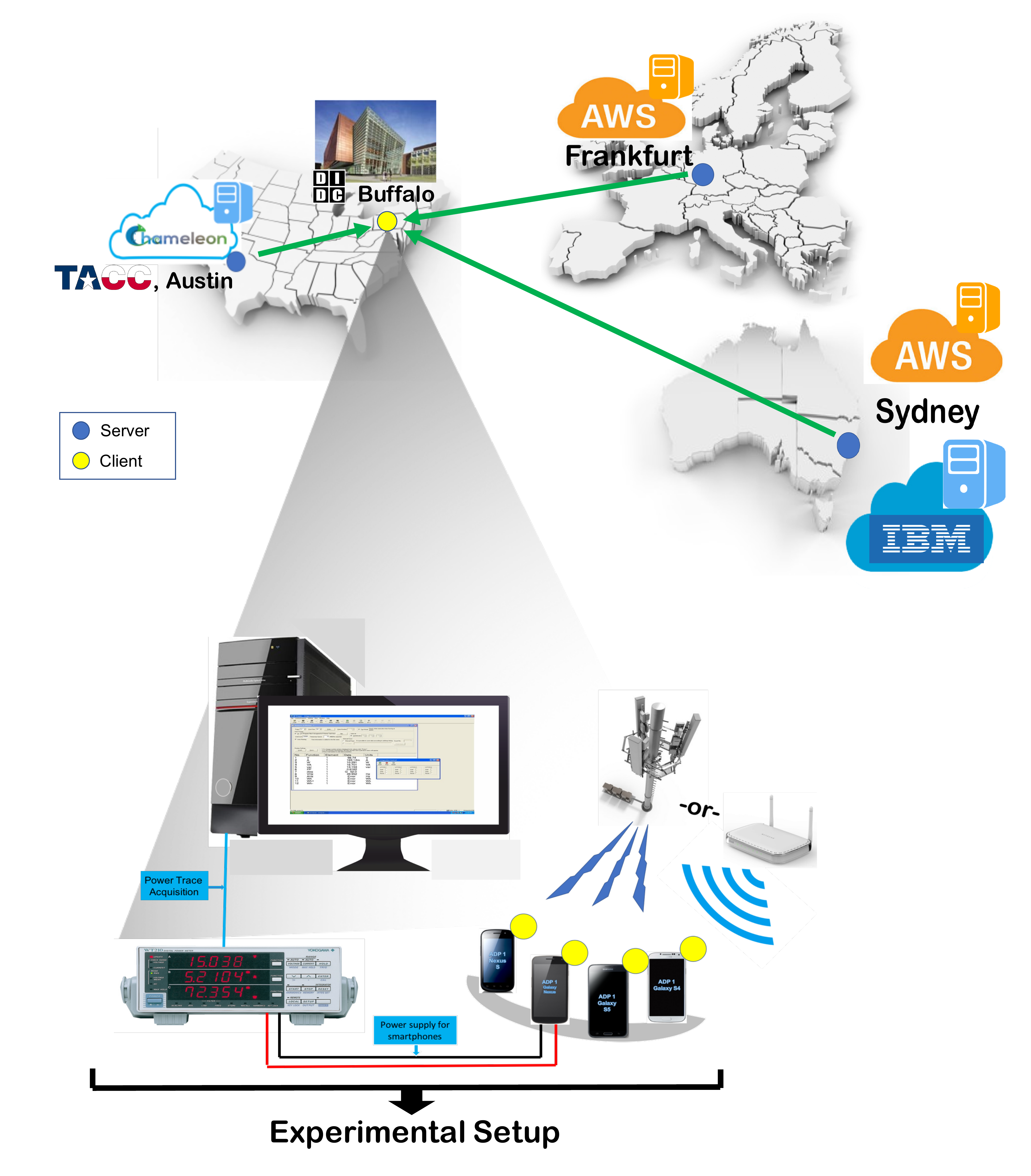}
		\vspace{-5mm}
		\caption{Network map of the experimental testbed and the setup of the power measurement system.}
		\label{fig:exSetup}
	\end{center}
	\vspace{-1cm}
\end{figure}

We analyzed the data transfer throughput of HTTP data transfers and the power consumption during which we run tests with different level of concurrency (cc), parallelism (p), I/O request size, and combined concurrency \& parallelism parameters. We also measured the instantaneous power consumption and total energy consumption of each individual request among different web servers and mobile clients.
The experiments were conducted on Amazon Elastic Compute Cloud (AWS EC2)~\cite{aws} instances, Chameleon Cloud~\cite{chameleon}, IBM Cloud ~\cite{IBM} and Data Intensive Distributed Computing Laboratory (DIDCLAB). The network map of the experimental testbed and the setup of the power measurement system are illustrated in Figure~\ref{fig:exSetup}.

In the experiments, we used six different types of files in order to analyze the effect of each individual parameter on transfer throughput and energy consumption. The details and characteristics of these files are presented in Table~\ref{tab:dataset}. 
In order to increase the robustness of the obtained throughput and energy consumption values for each experimental setting, we run each test within the range of five to ten times, and the average values of throughput and energy consumption were used. As a result of iteration of each individual experiment among four different mobile clients and three different web servers with different bandwidth (BW) and round-trip-time (RTT), we transferred varying size of nearly 1.8 Million individual files. Due to the space limitations of the paper, we had to limit the number of graphs we can present. The detailed analysis of the application-layer parameter effects on mobile data transfer performance and energy consumption are provided and discussed in the next section.

\section{Analysis of Parameter Effects}
\label{sec:experiments}

\setlength\belowcaptionskip{4ex}
\begin{figure*}[t]
	\begin{centering}
		\captionsetup{justification=centering}
		\begin{tabular}{cc}
			\includegraphics[keepaspectratio=true,angle=0,width=58mm]{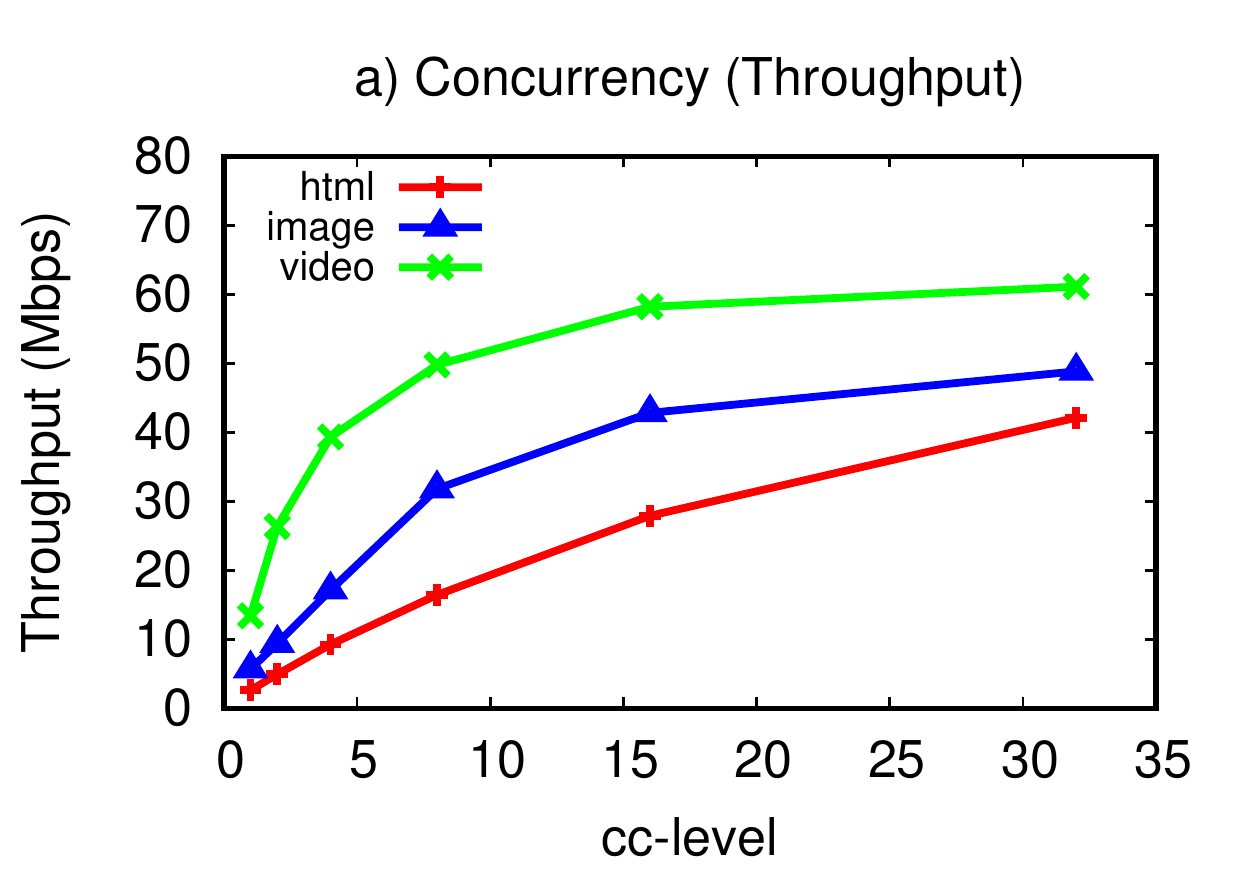}
			\includegraphics[keepaspectratio=true,angle=0,width=58mm]{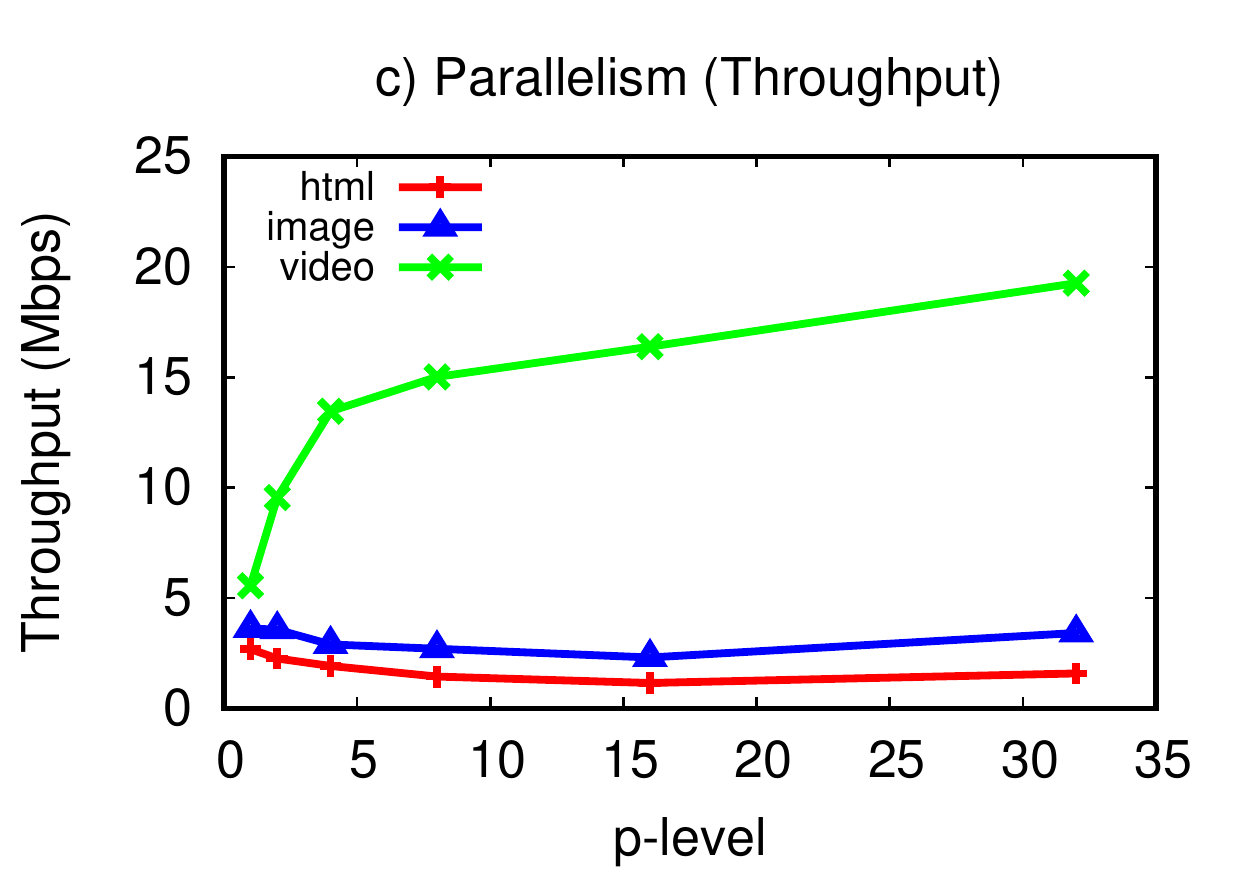}
			\includegraphics[keepaspectratio=true,angle=0,width=58mm]{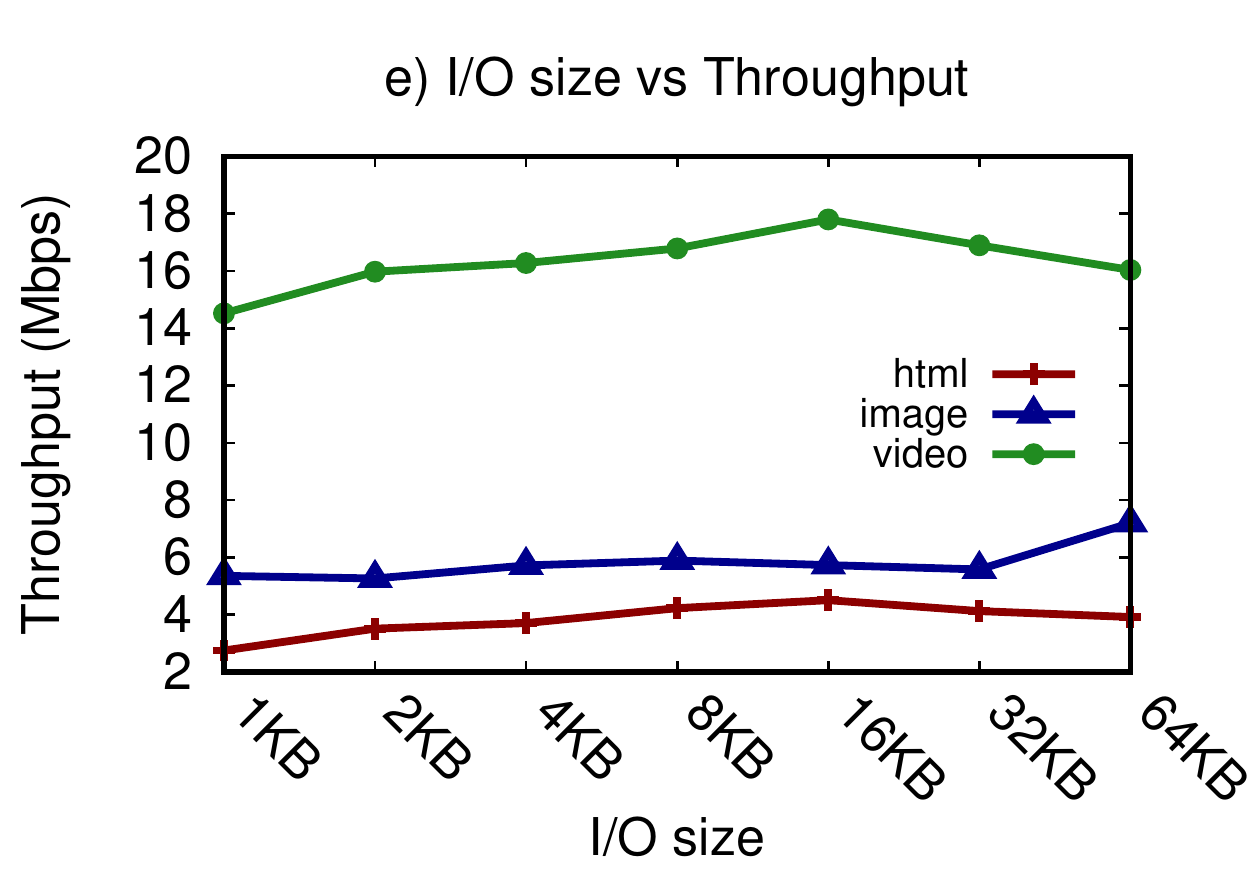}\\
			\includegraphics[keepaspectratio=true,angle=0,height=53mm,width=58mm]{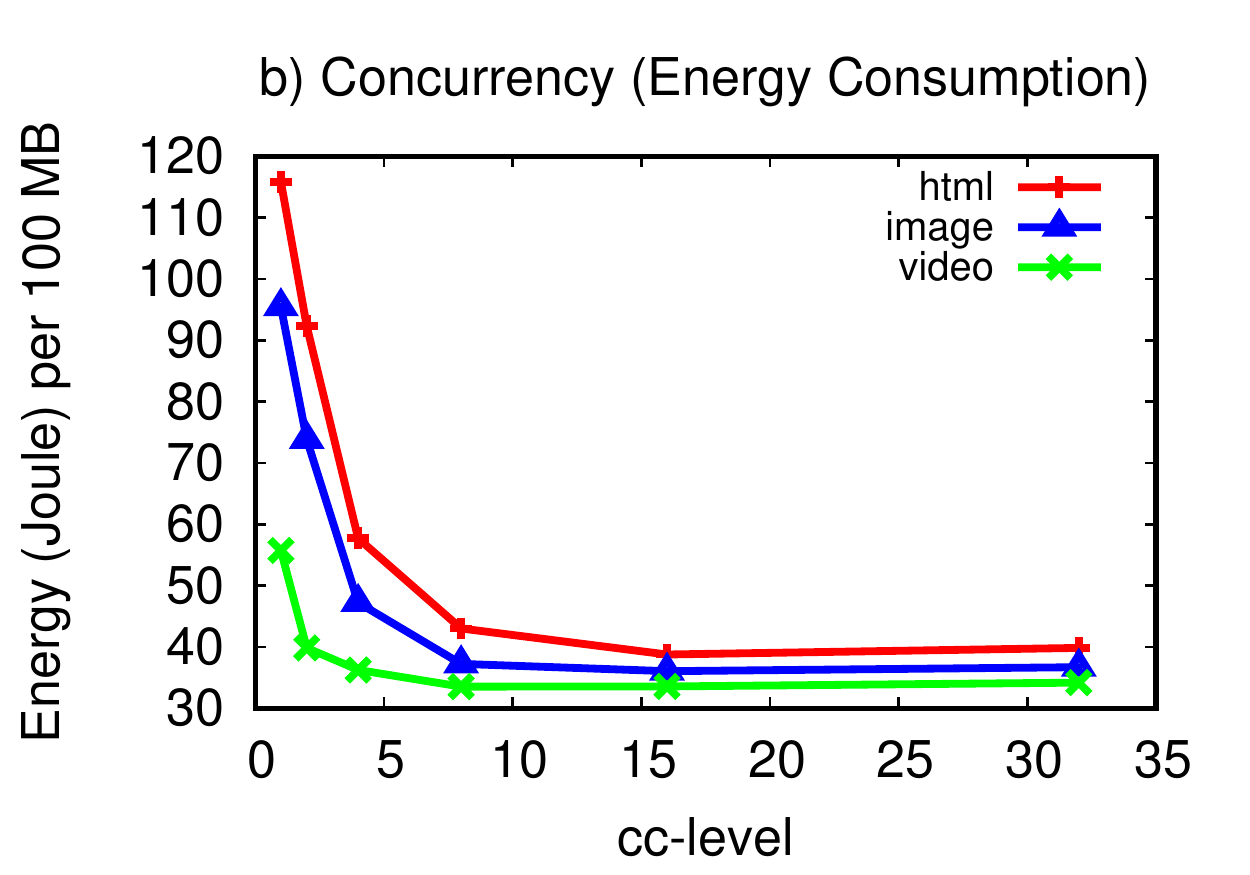}
			\includegraphics[keepaspectratio=true,angle=0,width=58mm]{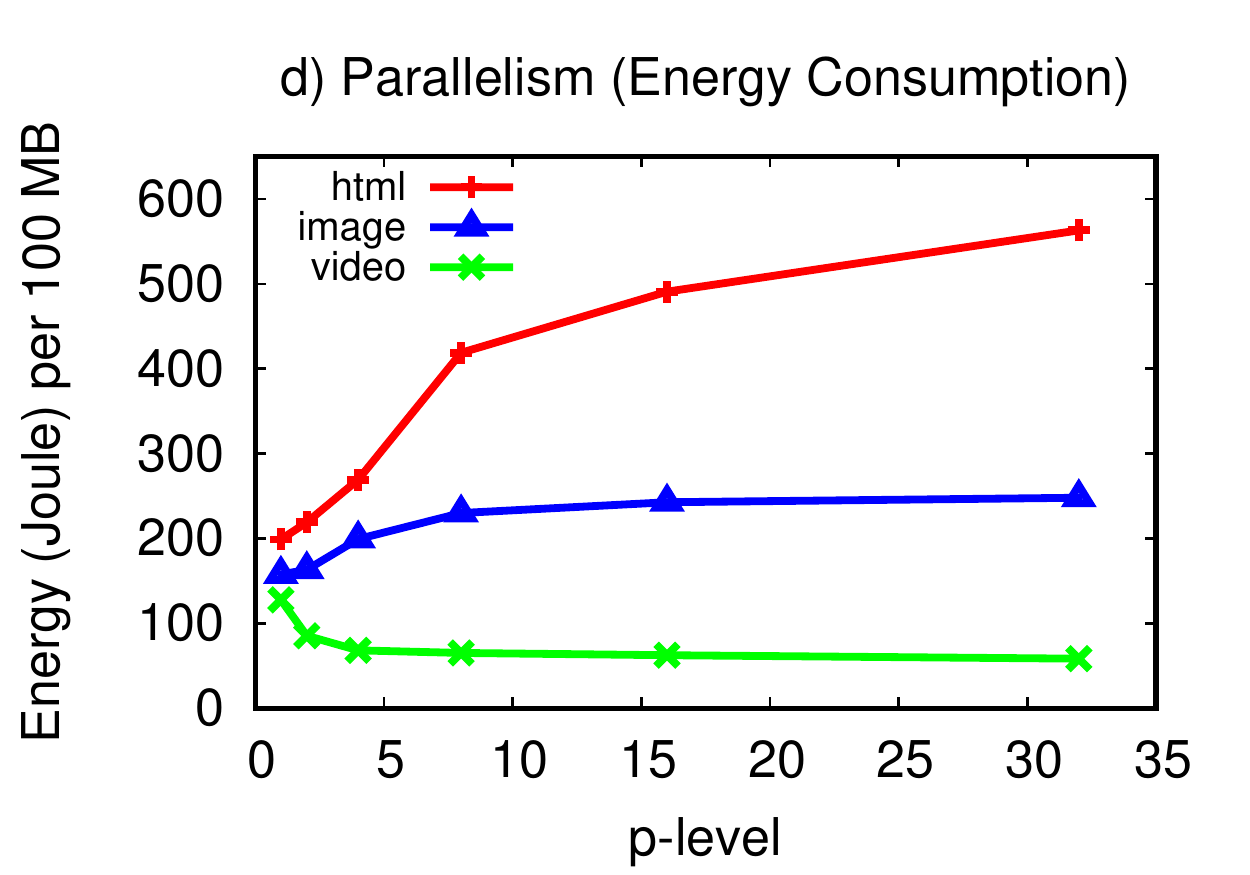}
			\includegraphics[keepaspectratio=true,angle=0,width=58mm]{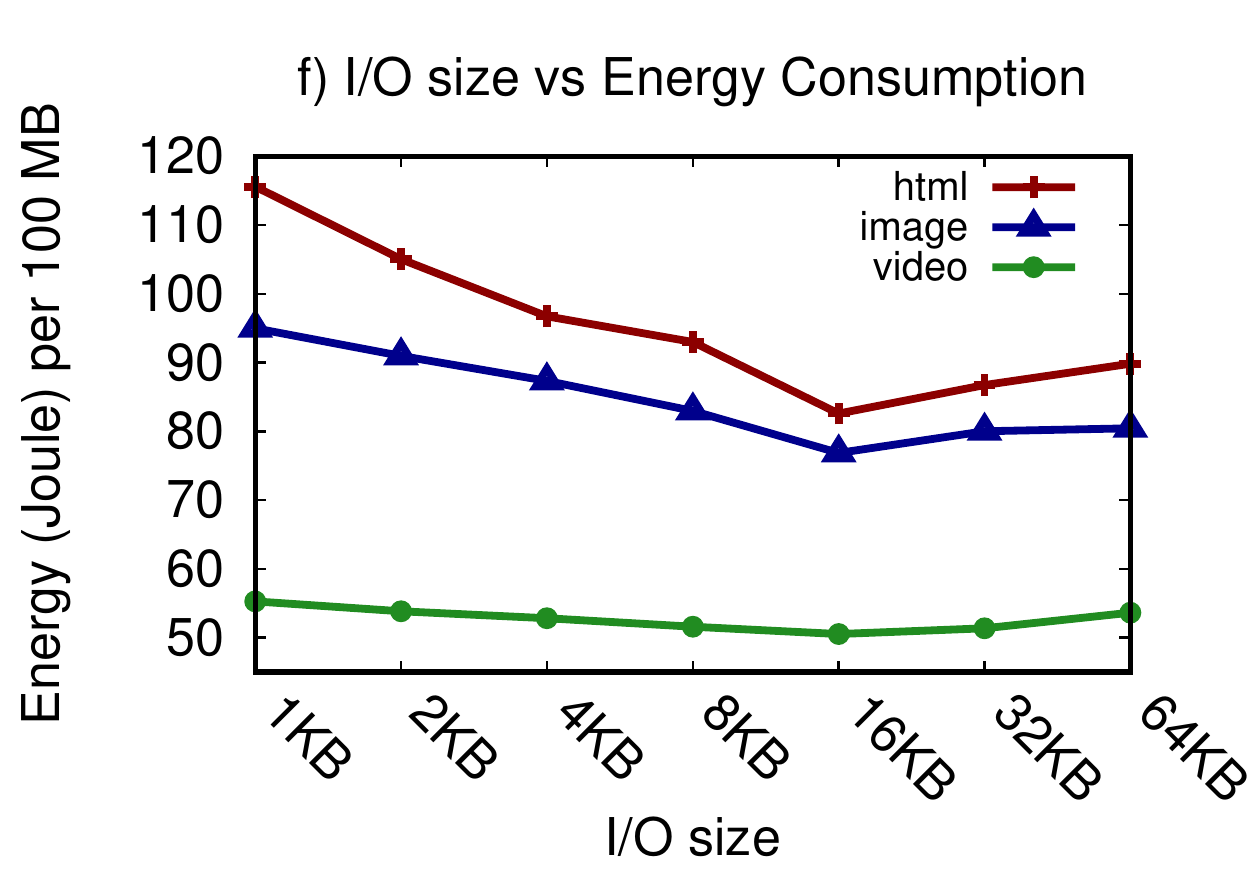}
		\end{tabular}
		\caption{Throughput vs Energy Consumption trade-offs of individual protocol parameters for WiFi data transfers between AWS EC2 Sydney and DIDCLAB Galaxy S5.} \label{fig:parameter}
	\end{centering}
	\vspace{-5mm}
\end{figure*}

We used four web servers at three different continents in our experiments. One of the web servers is located on the Chameleon Cloud at the Texas Advanced Computing Center (TACC) in Austin, Texas (USA). The other two servers are located at AWS EC2 in Sydney (Australia) and AWS EC2 in Frankfurt (Germany) respectively. These web servers are serving to the mobile clients at DIDCLAB in Buffalo, New York (USA). The tested mobile client devices include Samsung Galaxy S5, Samsung Galaxy S4, Samsung Galaxy Nexus N3 (also known as Galaxy Nexus L700), and Google Nexus S smartphones. 
We used Apache HTTP web server with default configuration and custom HTTP client configuration in our data transfer experiments. All experiments were controlled on the client side and the measurements were performed using a power meter as discussed in Section~\ref{sec:methodology}.

Figure~\ref{fig:parameter} presents the individual parameter effects of concurrency, parallelism, and I/O block size on the achieved throughput and energy consumption for the data transfers between the web server at AWS EC2 Sydney and the client Samsung Galaxy S5 at DIDCLAB in Buffalo. The energy consumption is measured per 100 MB of data transfers to achieve a standardized comparison among different data sizes. The RTT between the server and client is around 290 ms. Due to a shared network connection between the client and AWS EC2 Sydney server and limited WiFi bandwidth, the end-to-end transfer throughput of the video dataset increases from 13 Mbps up to 61 Mbps and then saturates. 

Overall, concurrency parameter showed a better performance than parallelism on our html, image and video datasets. When we increased level of concurrency from 1 to 32, it boosted end-to-end throughput for html, image and video datasets and reduced energy consumption on the smartphone client as seen in Figure~\ref{fig:parameter} (a)-(b). As we increase concurrency level from 1 to 16 for html dataset, the throughput almost doubled at each level and increased from 2.6 Mbps to 42 Mbps, which is 16.7X improvement. For the image dataset, the throughput increased from 5.6 Mbps to 48 Mbps (8.5X improvement), and for the video dataset it increased from 13 Mbps to 61 Mbps (4.5X improvement). Increasing the level of concurrency from 1 to 16 reduced total energy consumption 70\% for the html dataset, 68\% for the image dataset, and 38\% for the video dataset. After concurrency level 16, it still continued to keep energy saving balanced for html dataset while it started to increase the energy consumption for image and video datasets. On the other hand, when it comes to the parallelism parameter, the performance of each dataset showed different characteristics. Increased level of parallelism improved the end-to-end throughput of the video dataset transfers and decreased the energy consumption up to a specific level as shown at Figure~\ref{fig:parameter} (c)-(d). As the parallelism level increased, the throughput improved 3.1X and energy consumption decreased 45\%. On the other hand, parallelism did not improve the throughput for the html and image datasets. In fact, up to parallelism level 8, it gradually reduced the end-to-end throughput of html and image datasets and increased the energy consumption. Since these files are already small, partitioning them into smaller chunks and transferring them in multiple parallel streams did not help with filling the large network bandwidth.

There are other application-layer techniques for improving the end-to-end data transfer throughput and saving energy other than tuning network parameters (i.e., concurrency and parallelism) as well. Choosing optimal I/O request size of the application is a good example of one of these techniques as seen in Figure~\ref{fig:parameter} (e)-(f). With high speed networks, end-systems may become the bottleneck in terms of filling up these links, and sometimes a small tune up in application\textquotesingle s disk read/write speed would make a noticeable change in the overall end-to-end performance. 

\setlength\belowcaptionskip{4ex}
\begin{figure*}[t]
	\begin{centering}
		\captionsetup{justification=centering}
		\begin{tabular}{cc}
			\includegraphics[keepaspectratio=true,angle=0,width=65mm]{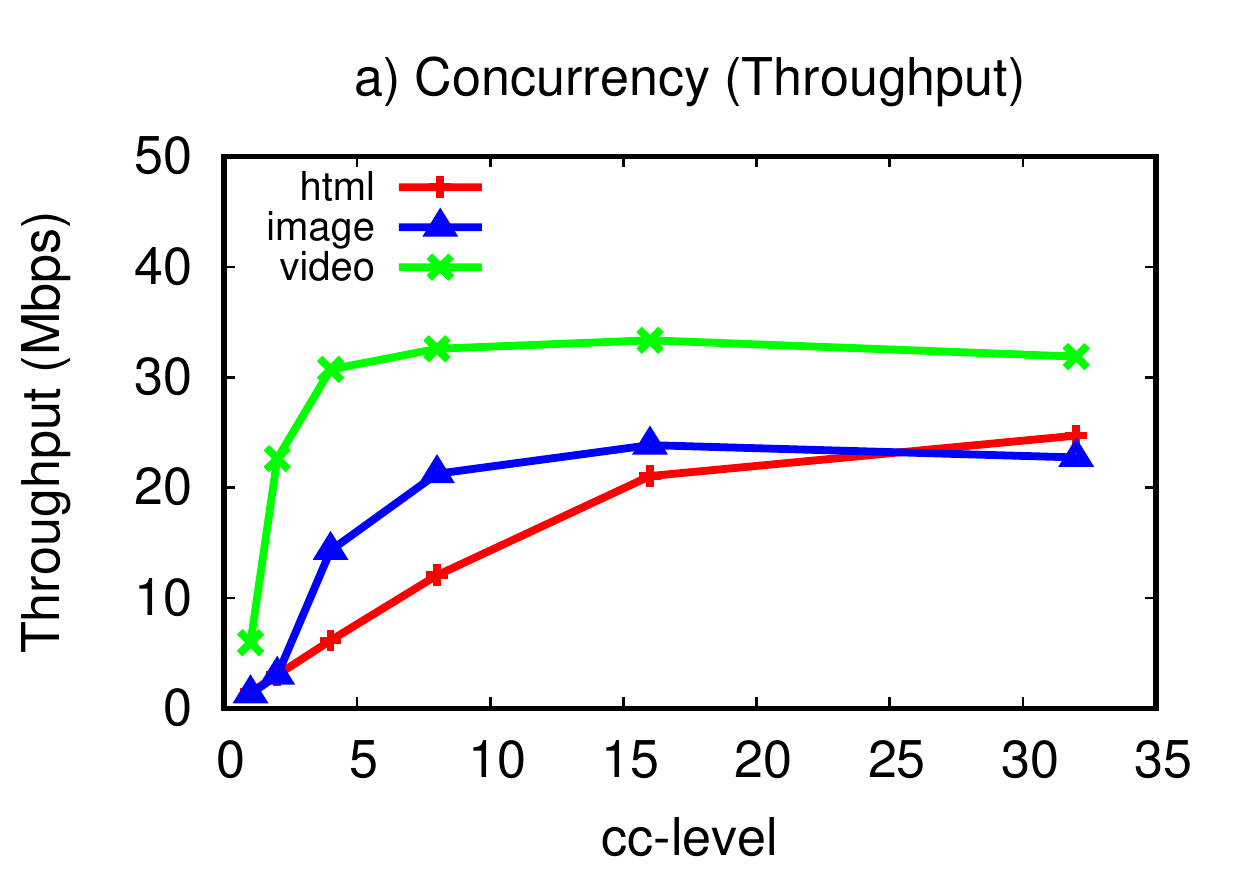}
						\hspace{1cm}
			\includegraphics[keepaspectratio=true,angle=0,width=65mm]{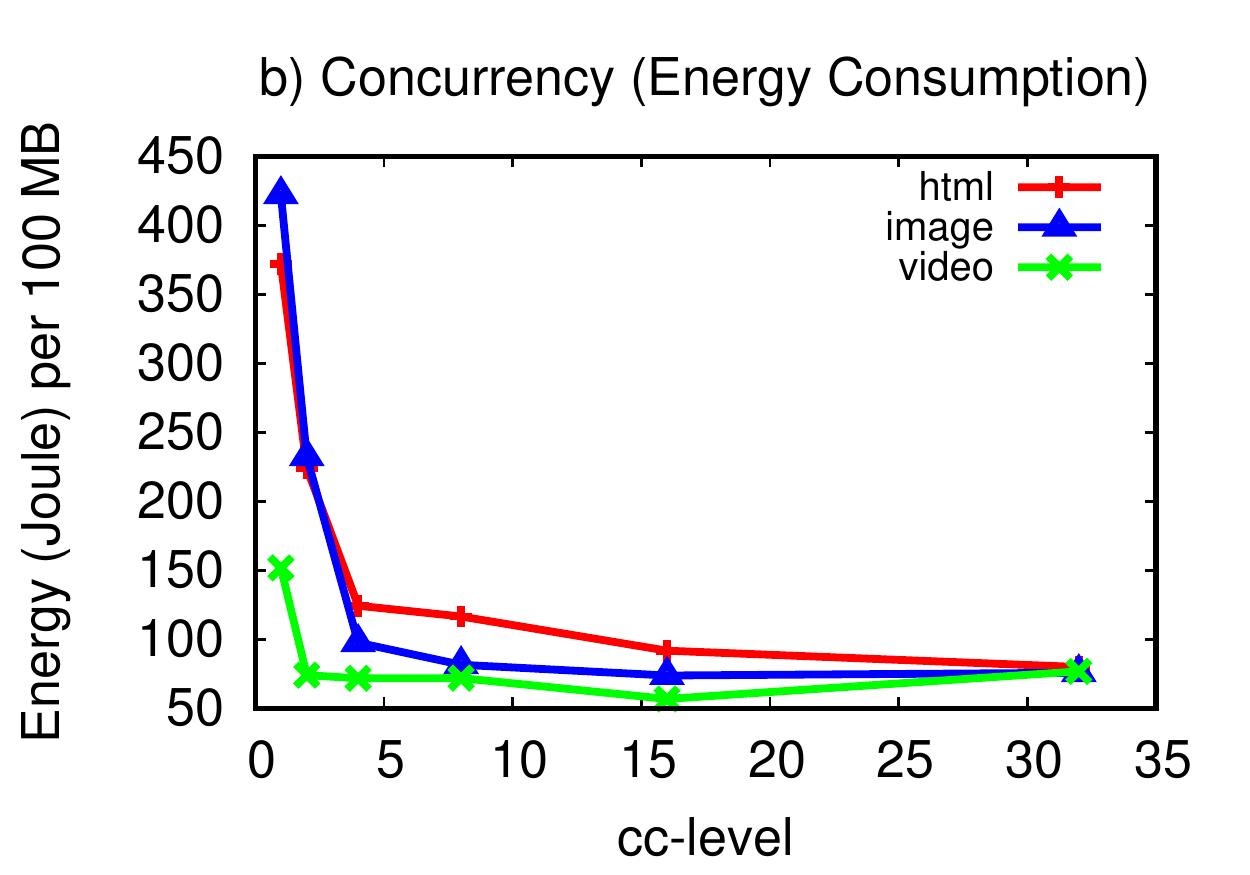}\\
			\includegraphics[keepaspectratio=true,angle=0,width=65mm]{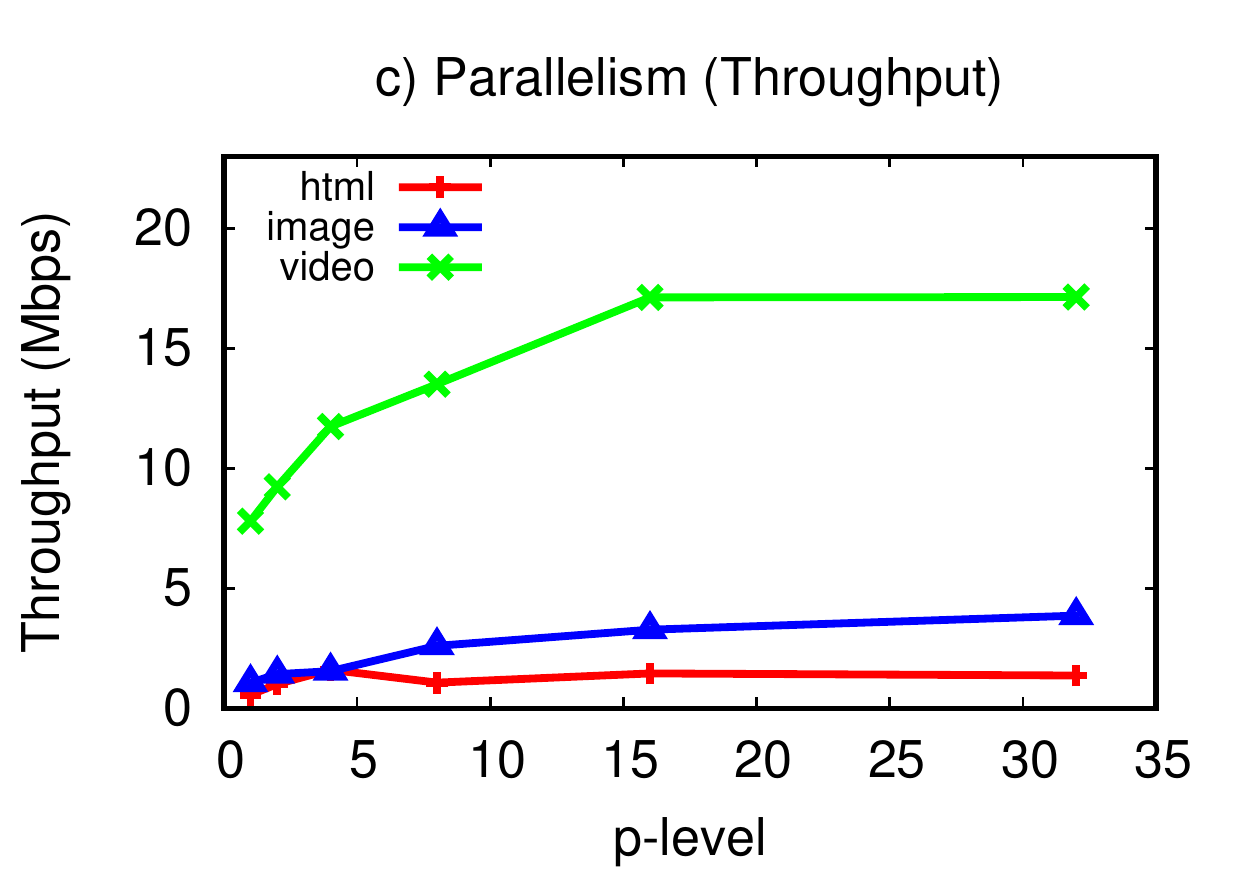}  
						\hspace{1cm}
			\includegraphics[keepaspectratio=true,angle=0,width=65mm]{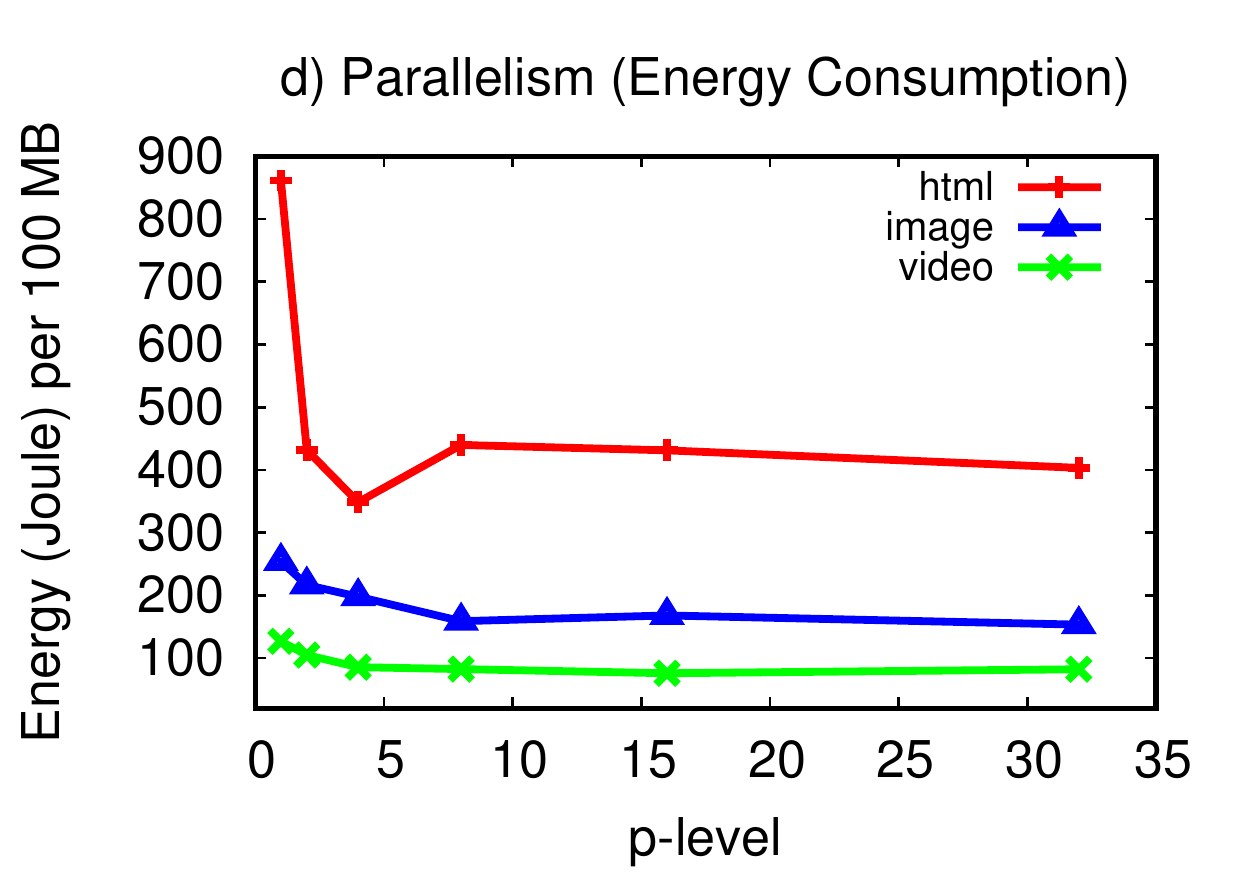}
		\end{tabular}
		\caption{Throughput vs Energy Consumption trade-offs of individual protocol parameters for 4G/LTE data transfers between AWS EC2 Sydney and DIDCLAB Galaxy S5.} \label{fig:4Gparameter}
	\end{centering}
	\vspace{-5mm}
\end{figure*}

After reaching optimal I/O request size of the application, further increase either does not change the already balanced system or causes a slight decrease. Figure~\ref{fig:parameter} (e)-(f) shows the change in throughput and energy consumption during video dataset transfers from AWS EC2 Sydney server to the client Galaxy S5 at DIDCLAB. We doubled I/O block size from 1 KB to 64 KB for all experiments. 
When I/O request size is increased from 1 KB to 8-16 KB, throughput slightly increased and came to a balance. Further increasing I/O request size induced increase in energy consumption at the same concurrency level, and it did not increase the throughput. The main reason for this is that the main bottleneck during the end-to-end data transfers in our testbed was not the mobile client's storage I/O speed, rather it was the wireless (WiFi or 4G LT) network connectivity.

Having throughput and energy consumption results of individual protocol parameters with WiFi connection, we also run the same individual parameter experiments with 4G LTE as presented in Figure~\ref{fig:4Gparameter}. Considering download/upload speeds of the cellular networks and WiFi, the results were quite similar, but with slightly lower end-to-end data transfer throughput. The noticeable difference was in the effect of parallelism on energy saving. Comparing the image and html dataset results with WiFi connection, increased level of parallelism had positively effected the energy saving with 4G LTE on these two datasets. This is mainly caused by the speed of the 4G LTE network. While the speed of 4G LTE network did not outdo achieved maximum throughput performance of concurrency parameter of WiFi connection, it increased the energy efficiency of html, image and video datasets. Comparing with WiFi connection, the energy saving increased from 70\% to 77\% for html, from 68\% to 76\% for image, and  from 38\% to 39.5\% for the video dataset. Again, increased level of concurrency became more effective for all datasets. Both parallelism and concurrency improved the end-to-end data transfer throughput and energy saving for all three datasets on 4G LTE data transfers.

\setlength\belowcaptionskip{5ex}
\begin{figure*}[t]
	\captionsetup{justification=centering}
	\begin{centering}
		\hspace*{-0.2cm}\begin{tabular}{ccc}
			\includegraphics[keepaspectratio=true,angle=0,width=58mm]{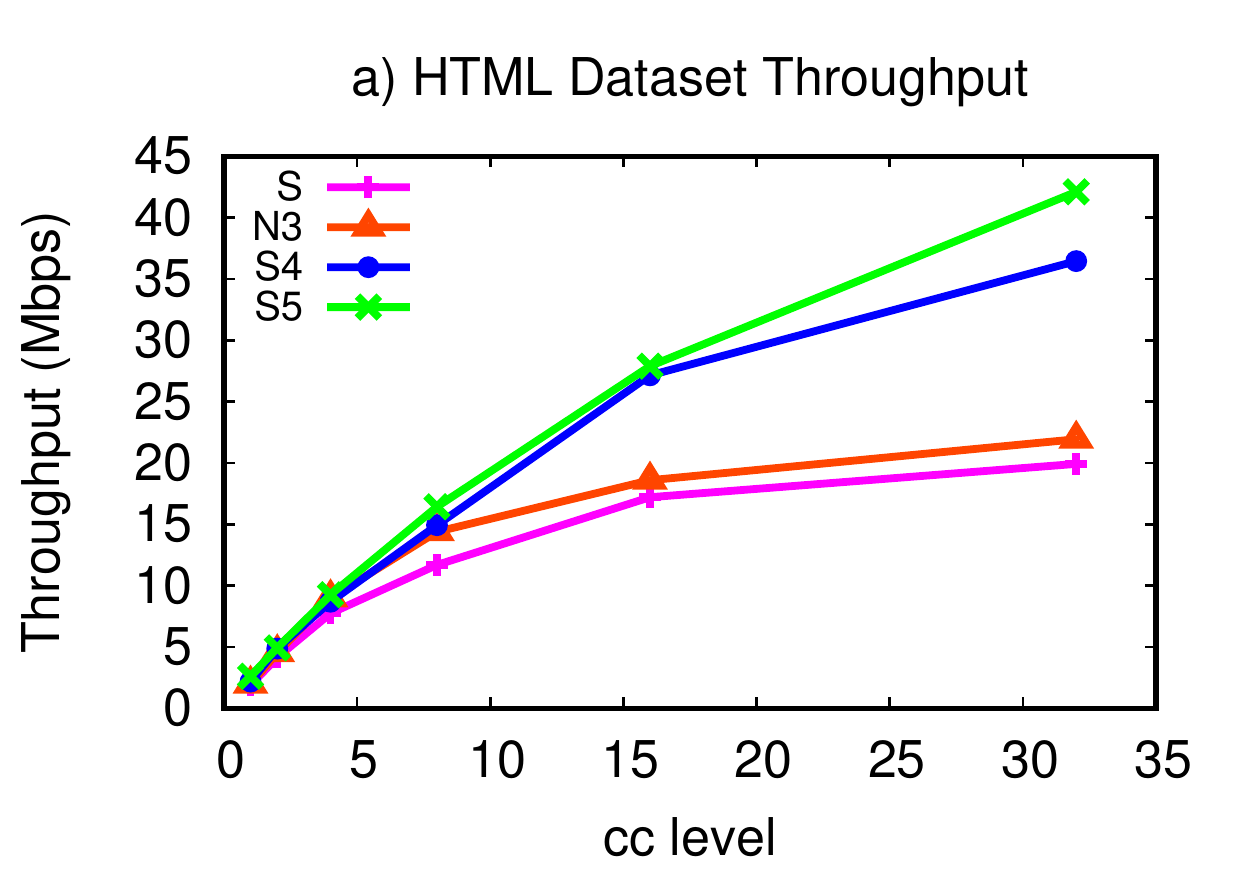}
			\includegraphics[keepaspectratio=true,angle=0,width=58mm]{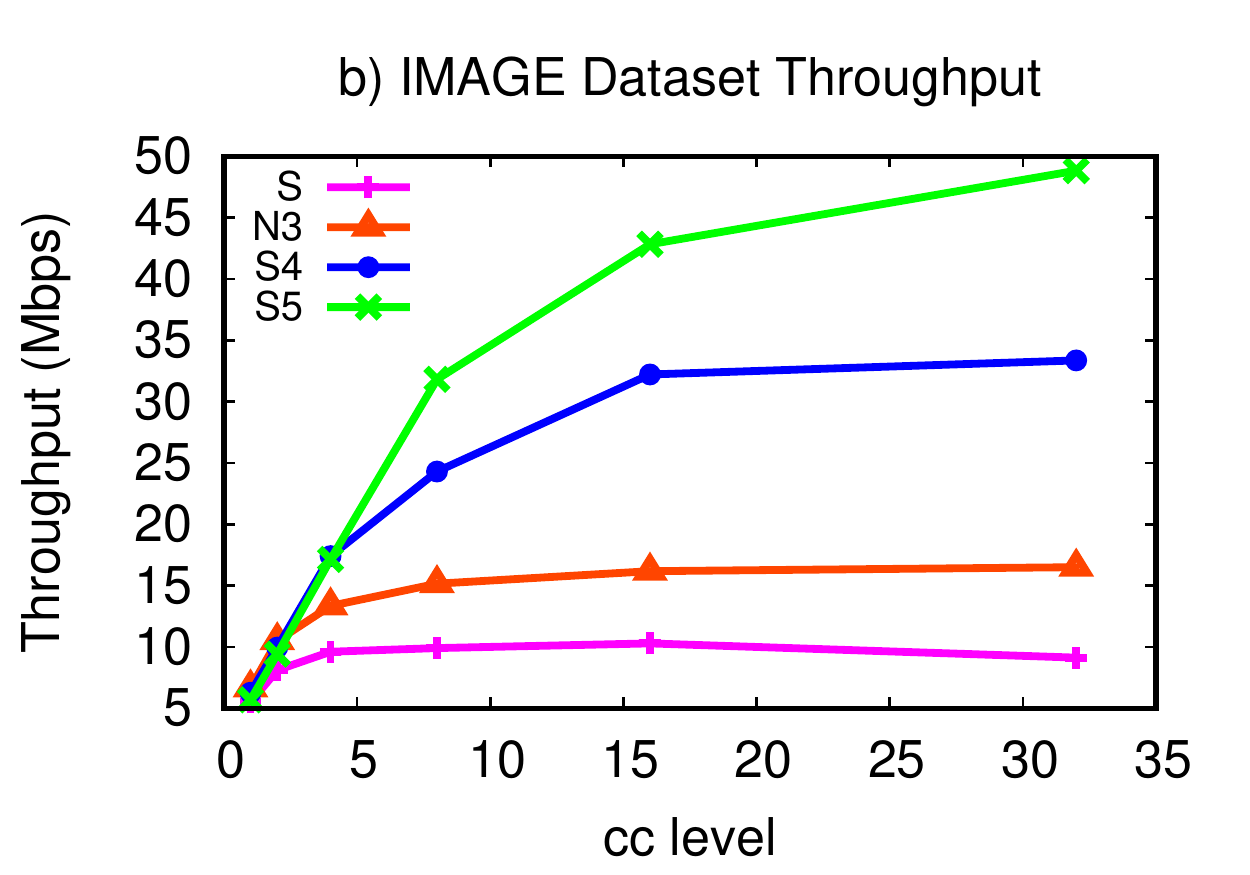}
			\includegraphics[keepaspectratio=true,angle=0,width=58mm]{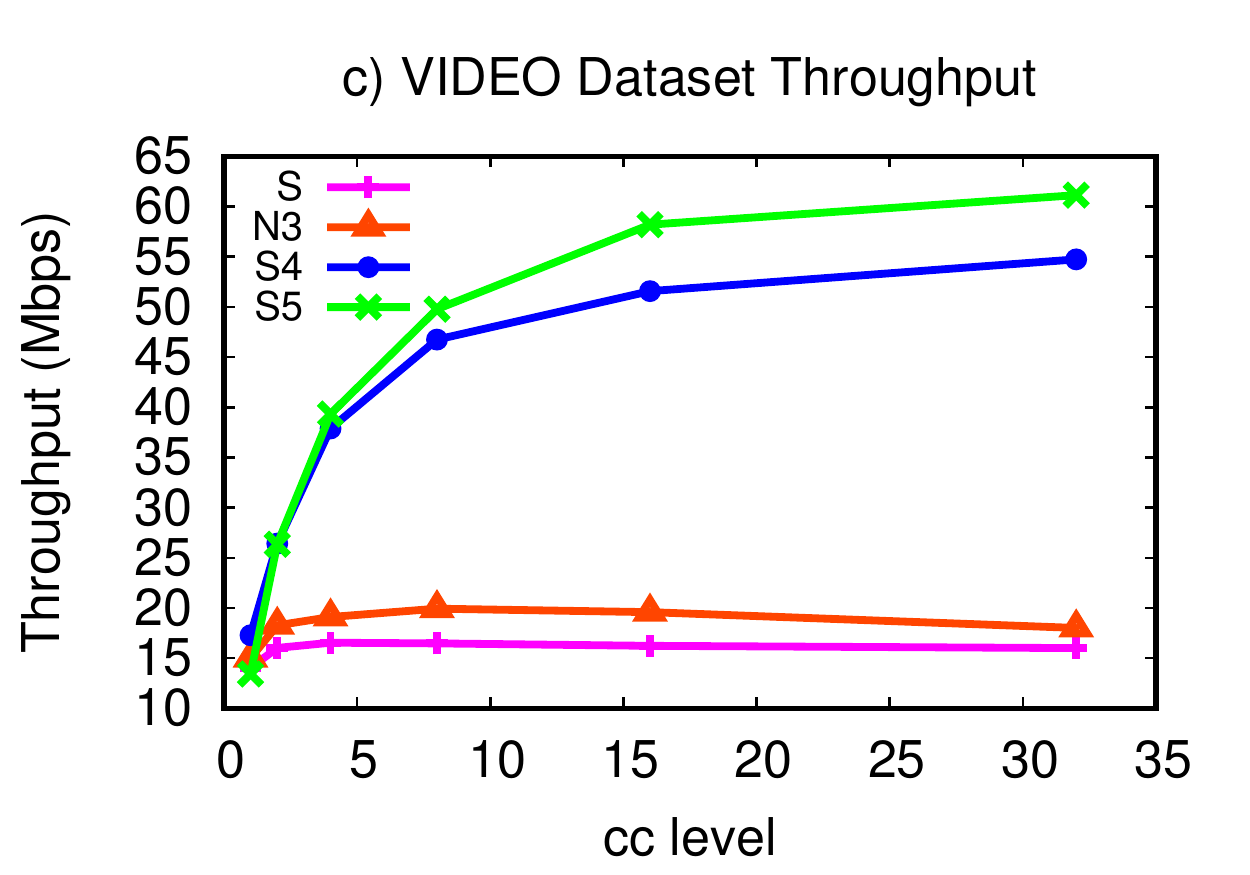}
			\\ 
			\includegraphics[keepaspectratio=true,angle=0,width=58mm]{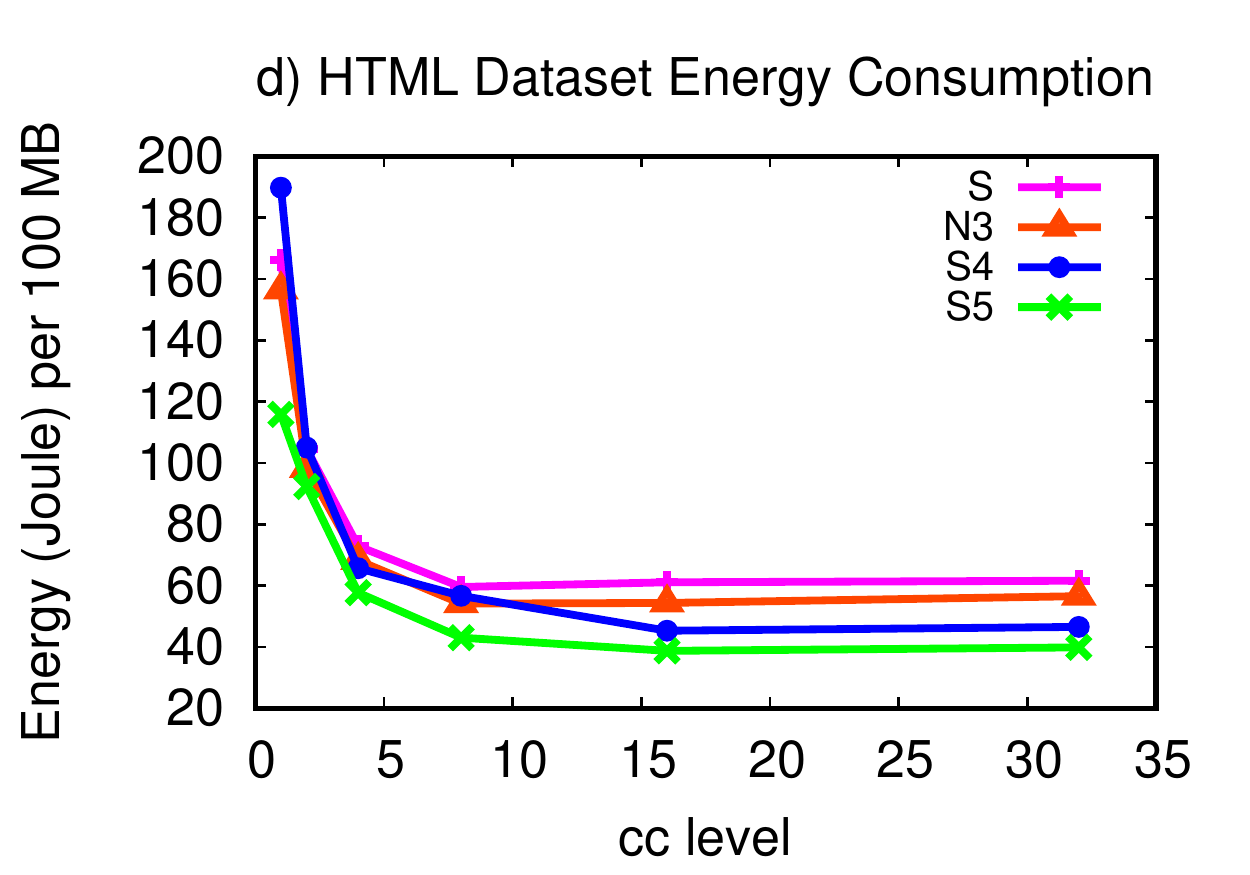} 
			\includegraphics[keepaspectratio=true,angle=0,width=58mm]{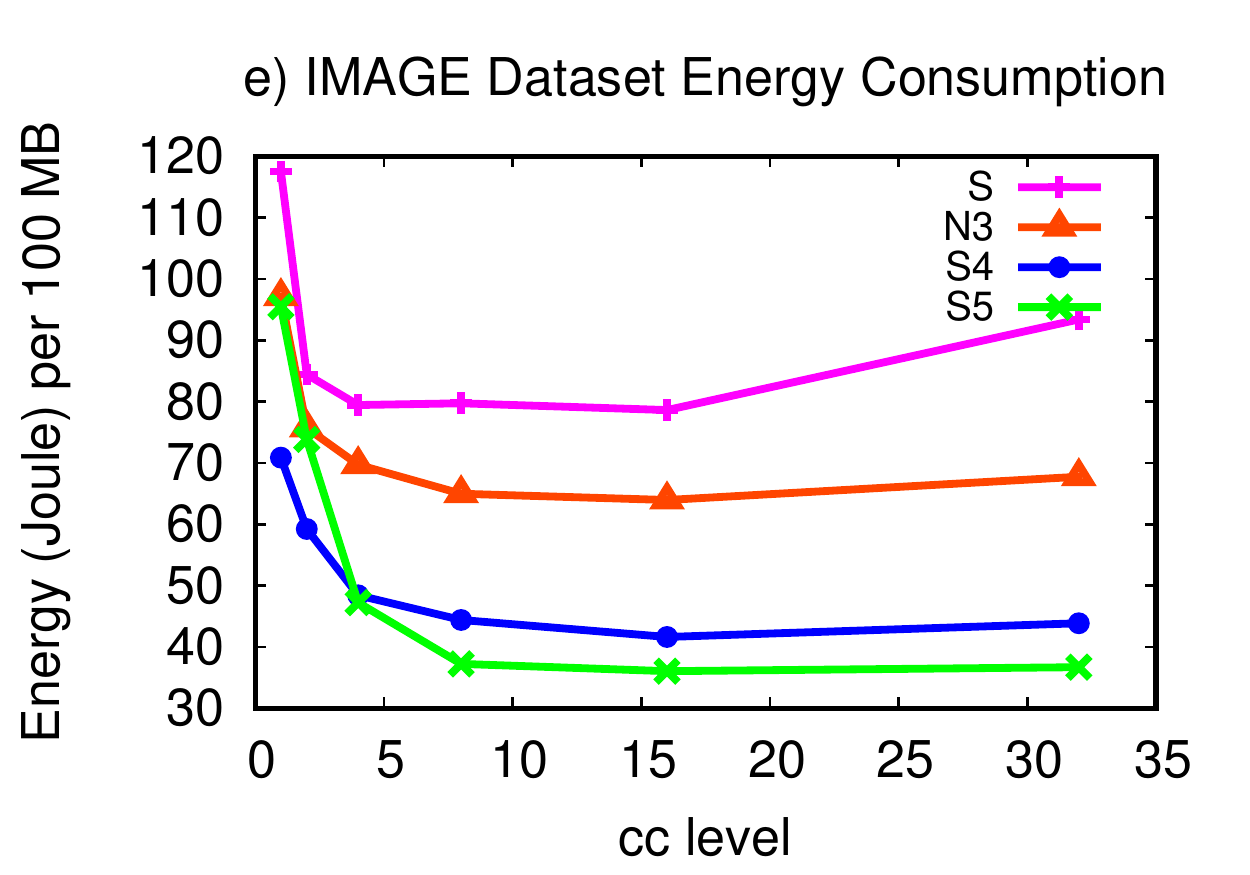}
			\includegraphics[keepaspectratio=true,angle=0,width=58mm]{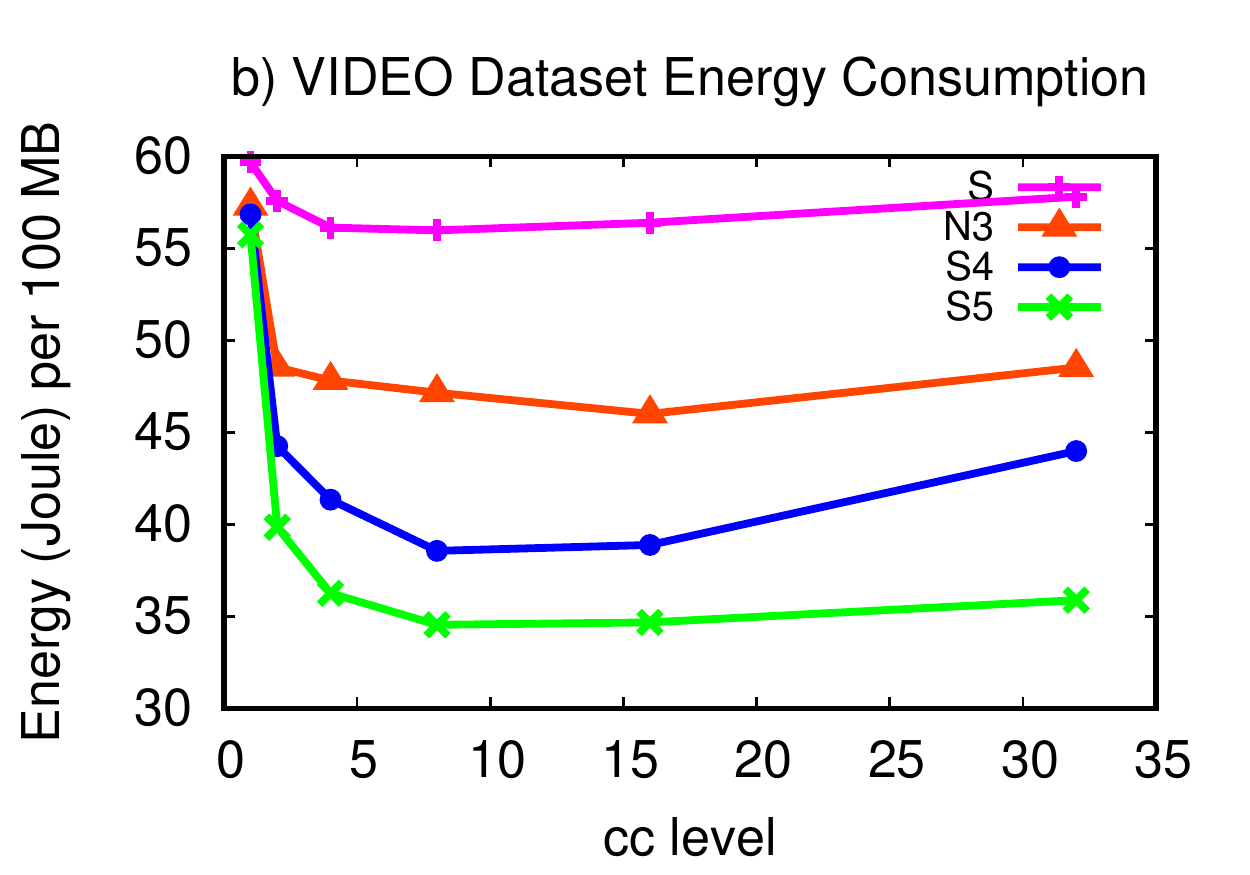}
\end{tabular}
	\vspace{-1mm}
	\caption{Throughput vs Energy Consumption per 100 MB data transfer  from AWS EC2 Sydney server to different mobile devices at DIDCLAB.} 		\label{fig:phones}
	\end{centering}
	\vspace{-6mm}
\end{figure*}

Next, we analyzed the impact of these parameters on different mobile devices. 
Figure~\ref{fig:phones} presents the results of achieved throughput and energy consumption for four different smartphone clients during data transfers from the web server on AWS EC2 Sydney. The specifications of the tested smartphones are
presented in Table~\ref{tab:phonespecs}. While earlier released smartphones such as Nexus S, Google Nexus N3 adopted mediocre throughput, recently released smartphones such as Galaxy S4, Galaxy S5 attained noticeably better performance. All phones showed improved end-to-end throughput for html, image and video datasets when concurrency level is increased up to a specific level. For html dataset,  concurrency parameter improved throughput for Nexus S by increasing throughput value from 1.8 Mbps to 19 Mbps (10.6X improvement), for Galaxy Nexus N3 by increasing from 1.9 Mbps to 22 Mbps (11X), for Galaxy S4 by increasing from 2.3 Mbps to 36 Mbps (15.6X) and lastly for Galaxy S5 by increasing from 2.6 Mbps to 42 Mbps (16.7X). For the image dataset, throughput increased 1.87X times for Nexus S, 2.4X for Nexus N3, 5.3X for Galaxy S4, and 8.5X for Galaxy S5. Lastly, throughput for video dataset was improved 1.14X times for Nexus S, 1.19X for Nexus N3, 3X for Galaxy S5 and 4.5X for Galaxy S5. Overall, best throughput was gained with Galaxy S5 using the html dataset, with almost 17X increase. 

\setlength\belowcaptionskip{4ex}
\begin{figure*}[!htb]
	\captionsetup{justification=centering}
    \begin{centering}
		\hspace*{-0.2cm}\begin{tabular}{ccc}
			\includegraphics[keepaspectratio=true,angle=0,width=58mm]{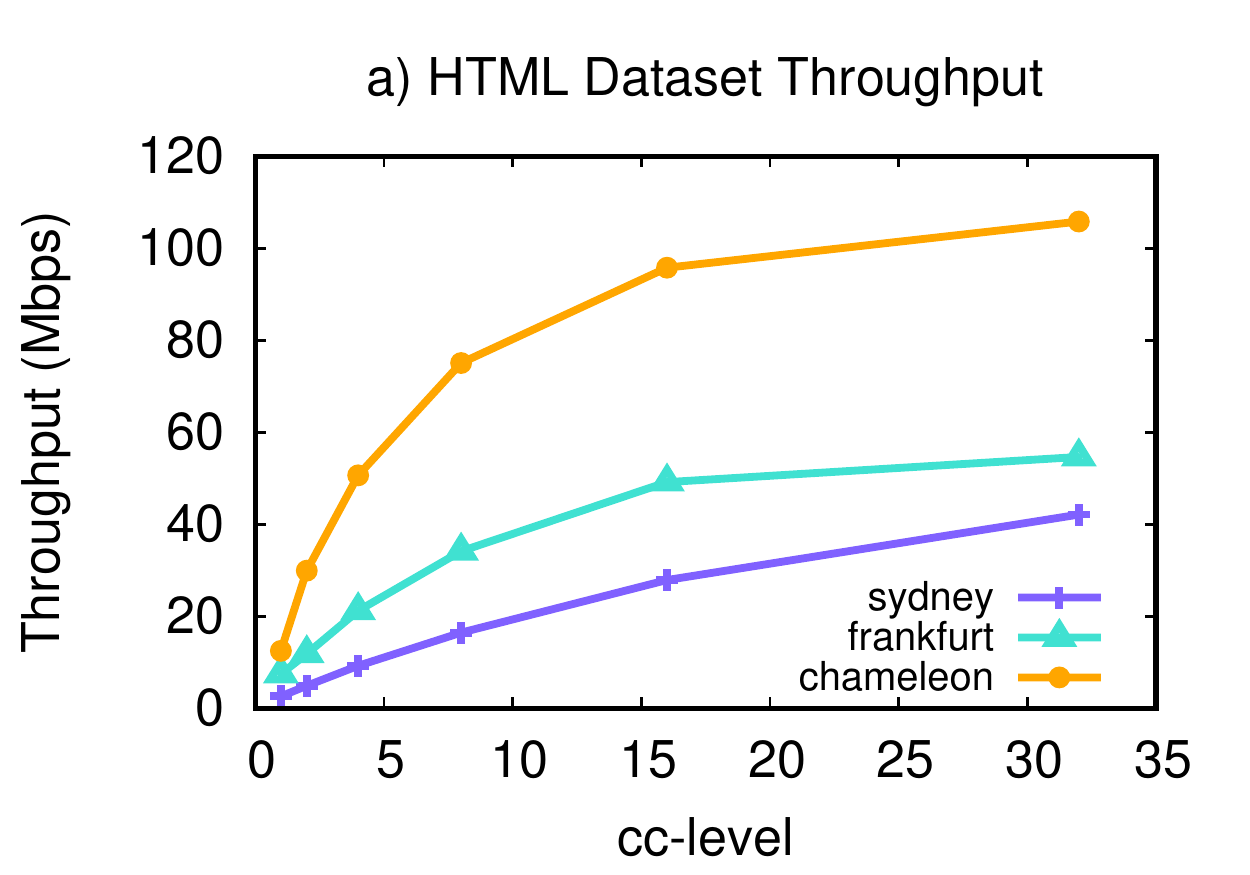}
			\includegraphics[keepaspectratio=true,angle=0,width=58mm]{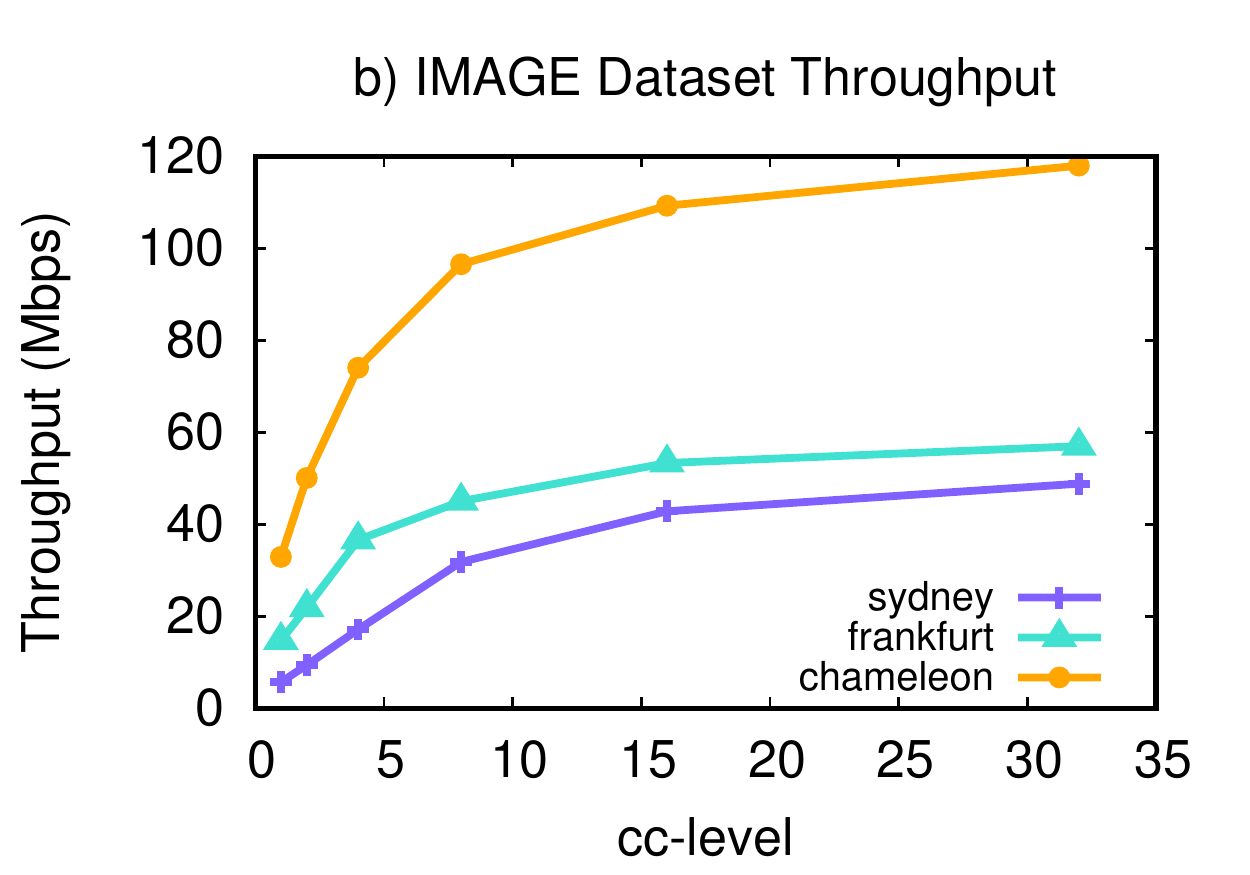}
			\includegraphics[keepaspectratio=true,angle=0,width=58mm]{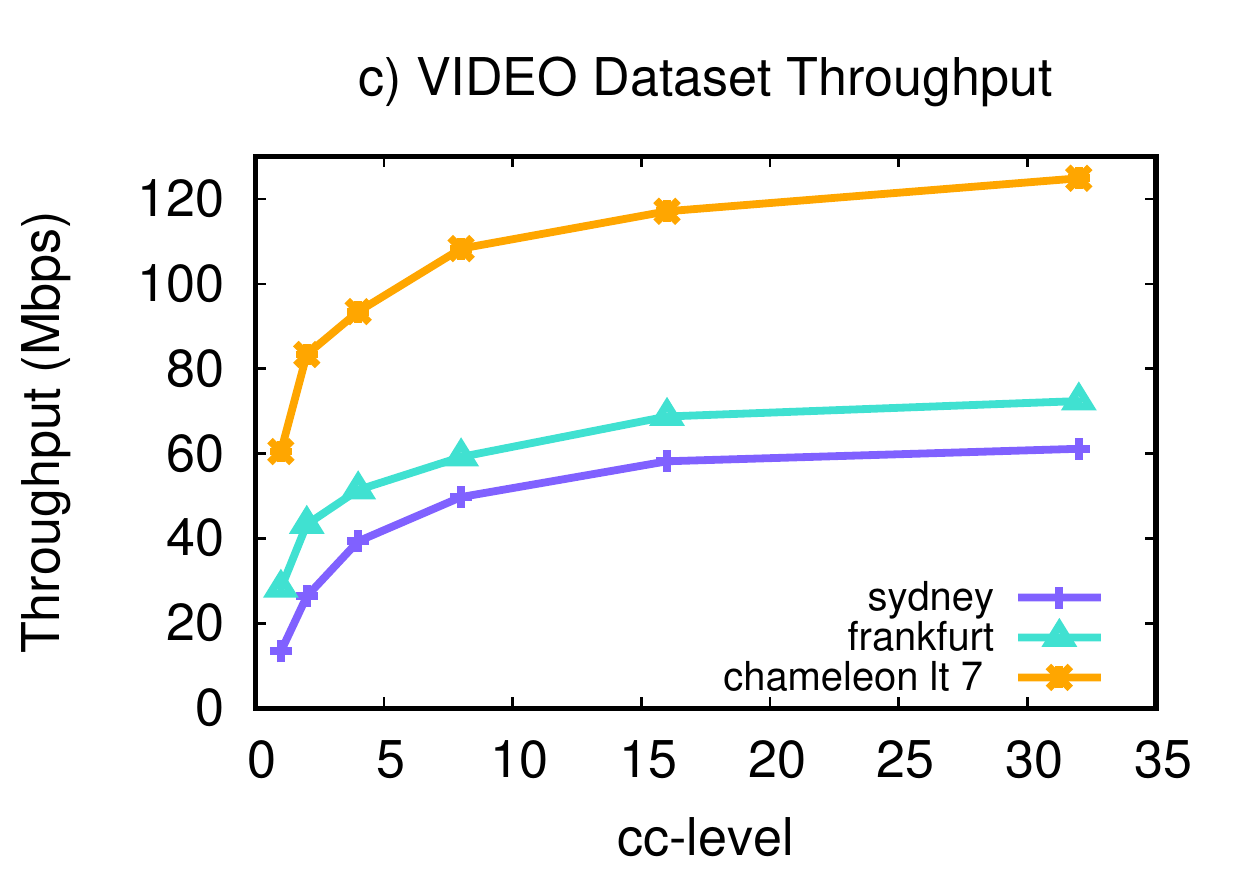}
			\\ 
			\includegraphics[keepaspectratio=true,angle=0,width=58mm]{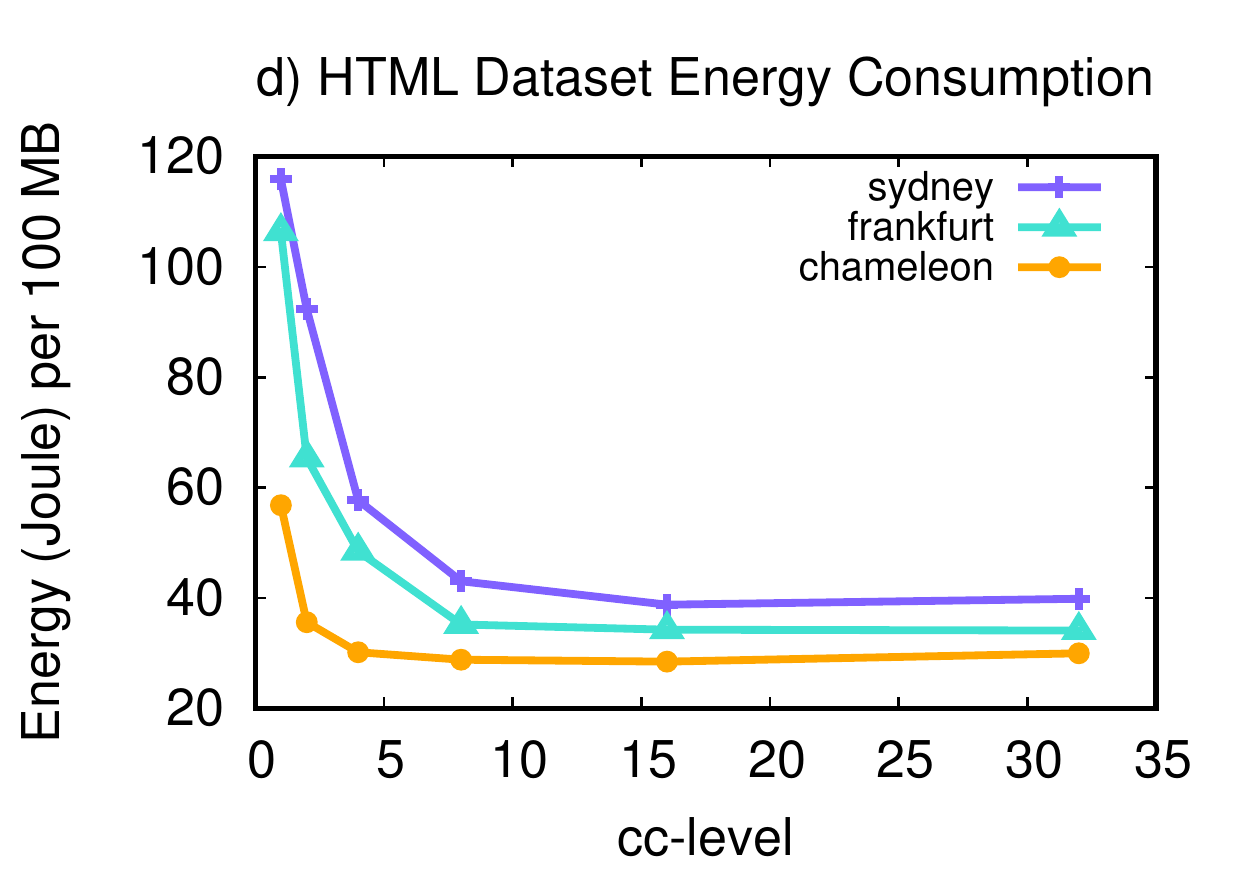} 
			\includegraphics[keepaspectratio=true,angle=0,width=58mm]{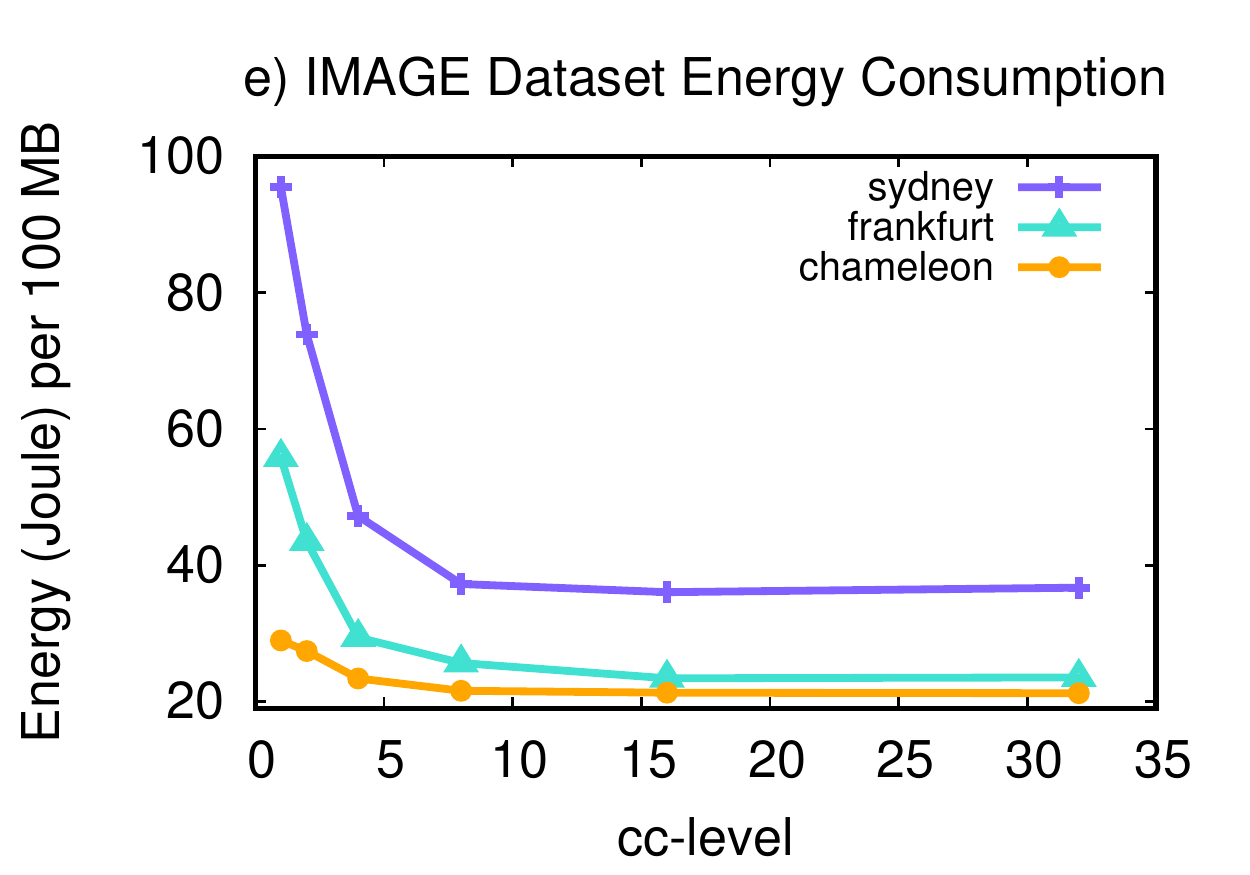}
			\includegraphics[keepaspectratio=true,angle=0,width=58mm]{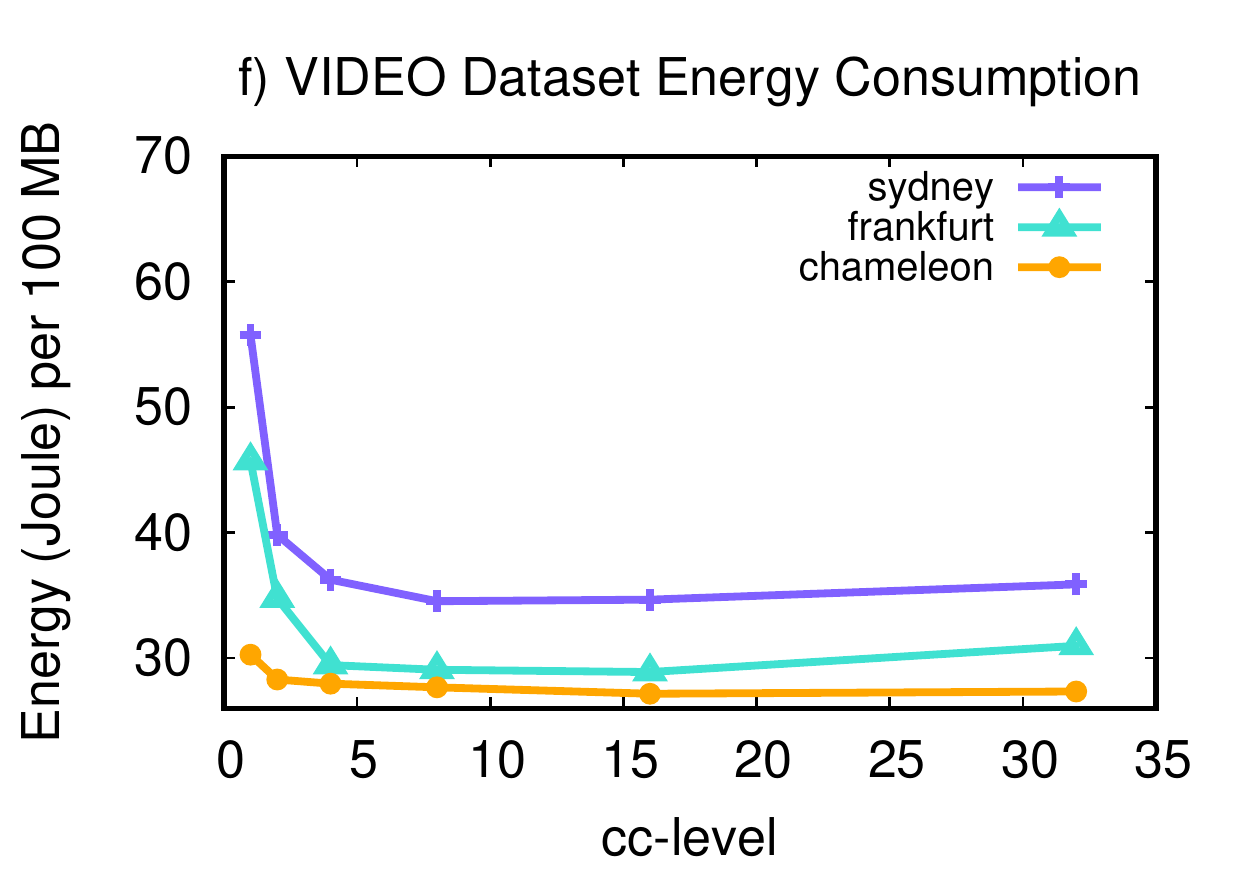}
		\end{tabular}
		\vspace{-1mm}
        \caption{Throughput vs Energy Consumption per 100 MB data transfer  from different servers to Samsung Galaxy S5 at DIDCLAB.} \label{fig:servers}
	\end{centering}
	\vspace{-6mm}
\end{figure*}

Each smartphone\textquotesingle s energy consumption per 100 MB also decreased as we increased concurrency level from 1 to 16 for the html dataset. While further increasing the level of concurrency caused an increase in the energy consumption for Nexus S and Nexus N3, it still continued to decrease for Galaxy S5, which obtained the highest energy saving with 70\% per 100 MB. The throughput of Nexus S and Galaxy Nexus N3 showed logarithmic behavior for the image and video datasets. In fact, concurrency doubled the throughput from 1 to 4 for image dataset and increased throughput by nearly 135\% for video dataset on Nexus S and Galaxy Nexus N3. Additionally, the energy saving rates of Nexus S and Galaxy Nexus N3 for the image dataset from concurrency level 1 to 4 were 33\% and 29\% and for video dataset 6\% and 9\%, respectively. On the other hand, we obtained higher throughput values for image and video datasets on Galaxy S4 and Galaxy S5. While throughput increased up to concurrency level 32, energy saving rate did not decrease as proportionally, namely it started to increase after concurrency level 16 for both S4 and S5. Galaxy S5 succeeded to increase the throughput 8.5X times and 4.5X times for image and video datasets respectively. It also saved 68\% energy for image dataset and 38\% for video dataset. The smartphones with powerful processors, larger memory and optimized OS that is high compatible with all subsystems distinctively separated from earlier released ones. 

We also analyzed the impact of the web server location and conducted experiments from three different web servers in Austin, Frankfurt, and Sydney to the mobile client at DIDCLAB in Buffalo. 
The RTT between mobile client at DIDCLAB to Chameleon Cloud (Austin, TX), AWS EC2 Frankfurt, and AWS ECS Sydney are 59 ms, 115 ms, and 290 ms respectively. The average throughput vs energy consumption trade-offs from those three servers to the client Samsung Galaxy S5  at DIDCLAB can be seen in Figure~\ref{fig:servers}. As shown in Figure~\ref{fig:servers}(a)-(f), increasing the level of concurrency up to 16 improved the end-to-end data transfer throughput while reducing the energy consumption. 
Although the throughput curves showed similar increasing pattern for all three servers, the highest throughput is achieved from the Chameleon Cloud in Austin node, which is 105 Mbps for the html, 117 Mbps for the image, and 124 Mbps for the video dataset. Energy consumption rates on three servers also showed similarity in decreasing when the concurrency level is changed from 1 to 16, and then it comes to a balance. Overall, we obtained the best energy savings for the html dataset, which was 50\% on Chameleon Cloud, 70\% on EC2 Sydney, and 65\% on EC2 Frankfurt. 

\section{Proposed Model}
\label{sec:proposedmodel}

\begin{figure*}[t]
\begin{centering}
\includegraphics[keepaspectratio=true,angle=0,width=140mm]{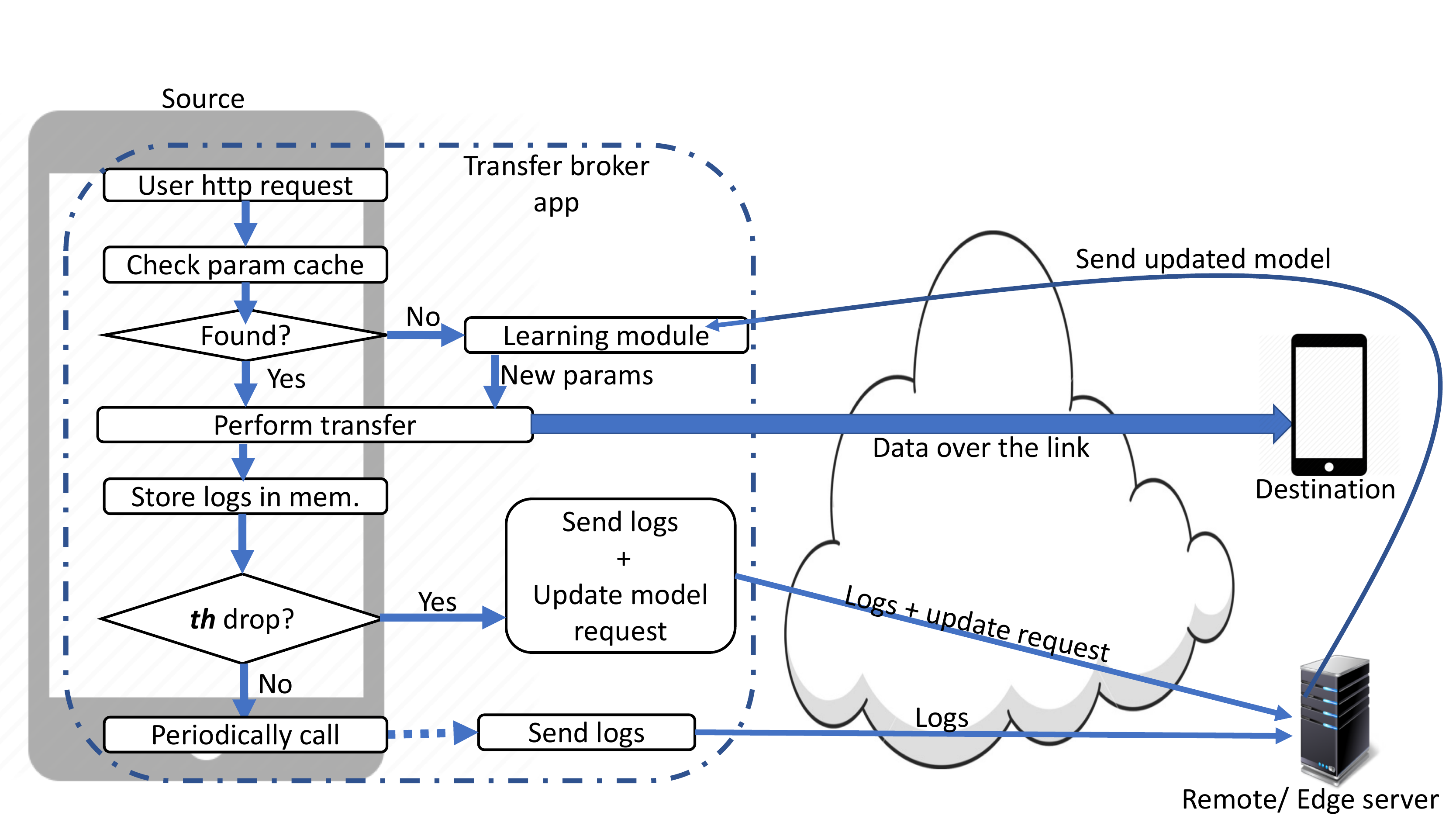}
	\end{centering}
    \vspace{-3mm}
\caption{Overview of the FastHLA model.} 
\label{fig:proposed_model}
\vspace{-3mm}
\end{figure*} 

Dynamic nature of the real-time background traffic on network links has a profound impact on the data transfer performance, and makes it very challenging to predict the optimal parameter combination to achieve the highest throughput possible. 
Historical data transfer logs can provide useful information about the data transfer pattern on a given link and the achievable throughput behavior. 
We define optimization based on this approach as Historical Log Analysis (HLA). However, the only historical analysis is not enough to keep up with the dynamic network conditions. We also need the current network status to decide and dynamically tune the parameter settings. We define this approach as Online Network Probing (ONP). An ideal solution might be combining both approaches to find the best parameter settings for the transfers. HLA model is an offline analysis model that takes historical transfer logs as input and finds optimal parameter settings for the requested data transfer.  
However, We have to take into account several design challenges explained below.

\theoremstyle{remark}
\newtheorem{challenge}{Challenge}
\begin{challenge}\label{ch:1}
Mobile devices are not suitable for compute-intensive historical analysis. Historical analysis needs to be done outside the mobile device. 
\end{challenge}

\begin{challenge}\label{ch:2}
Historical analysis introduces a new cost on both computation and energy consumption. We also have to consider the frequency of the historical analysis as each full iteration of the analysis will introduce more compute and power cost.  
\end{challenge}

\begin{challenge}\label{ch:3}
The benefit of figuring out the optimal parameter setting has to outweigh a transfer without any optimization. Assuming $C$ is the cost function, we can strictly constrain $C$ as follows -  
\begin{equation}
C(HLA) + C(T_{opt}) < C(T_{no{\text -}opt})
\end{equation}
Here, $T_{opt}$ is the transfer with optimized parameters and $T_{no{\text -}opt}$ is the transfer without any optimization. 
\end{challenge}

\begin{challenge}\label{ch:4}
Mobile devices have limited memory, therefore, we should allocate a fixed memory size to store the logs. Due to the fixed memory size, it is possible that the new logs can overwrite the old historical logs. 
\end{challenge}

\begin{challenge}\label{ch:5}
We also need to make sure that the communication between the historical analysis server and source device is minimal. Too much communication can take toll on the data transfer throughput.
\end{challenge}

To address these challenges, we have introduced a dynamic framework called \textit{FastHLA} as presented in the next subsection.

\subsection{Fast Historical Log Analysis (FastHLA)}
Conceptually, FastHLA outsources the analysis of historical data transfer logs to an edge server or to the cloud. The historical analysis will introduce additional computation and energy consumption. Even if we outsource the task it is still consuming computational resources and power on the edge server or in the cloud. \textit{Challenge}~\ref{ch:3} might seem counter-intuitive at the beginning, however, we have seen that a transfer with sub-optimal parameter choice achieves low throughput which leads to longer transfer time and high power consumption. We designed FastHLA in a way that it does not need to be run for every transfer. Therefore, the cost of FastHLA can be amortized over many subsequent transfers. To generalize the model even farther we can run FastHLA for many mobile devices in the cloud to amortize the FastHLA cost over many mobile devices. 

An overview of the FastHLA model is provided in Figure~\ref{fig:proposed_model}.  We introduced a light-weight transfer broker that receives a transfer request and performs transfer with best possible parameters. Network condition does not change significantly over a short period of time, however, when it changes the previous optimal choice of parameters might become sub-optimal. Therefore, running FastHLA once in the beginning is no better than statically setting the transfer parameters (an approach used in many current solutions). Therefore, we need a strategy to minimize the frequency of running FastHLA. To resolve this issue, we introduced a caching mechanism for previous optimal parameters and a Learning Module (LM) in the mobile device. The parameter cache is a dictionary structure that maps network condition to parameter value list. On the other hand, the LM can take user request and network condition as input and provide best known parameters. The training of the LM is performed as a part of FastHLA outside the mobile device. Each time the FastHLA runs, it updates the local Learning Module. A trained LM in the mobile device can provide predicted optimal parameters in almost constant time. Transfer broker first looks into the cache for the parameter settings. In case of a cache miss, it asks parameters from the LM and performs the transfer using those parameters. It is crucial to keep LM up-to-date. It can be done by accepting periodic updates from FastHLA. However, there might be a highly unlikely case where the network condition is unknown to the LM itself and the parameters provided by LM are sub-optimal. In that case, an immediate update request will be issued only if there is a significant drop in data transfer performance. However, the chance of such miss significantly reduces after each FastHLA update, because the training is an additive process with proper generalization method. Therefore, LM gets more and more precise after each FastHLA update. 

To address memory issue explained in \textit{Challenge}~\ref{ch:4}, the transfer broker periodically sends historical transfer logs to the remote server where HLA is performed, so that old historical logs become available to HLA before being overwritten by the new logs. The communication overhead explained in \textit{Challenge}~\ref{ch:5} are the periodic update requests and periodic log transfers. We do not need to communicate with HLA server during the transfer except during the highly unlikely case explained above where both cache and LM fails to provide optimal parameters. To reduce the communication overhead during the transfer, we decided not to include the Online Network Probing into our model. We can use the most recent logs to assess the network condition.  

The model consists of five steps - 
(1) historical transfer log collection and preprocessing; 
(2) clustering similar logs; 
(3) optimization;
(4) learning optimal parameters;
and (5) scheduling mixed sized data. The details of these steps are explained below.

\vspace{1mm}
\noindent {\bf Step 1 -- Historical log collection and preprocessing:} 
We collect historical logs for the previous data transfers. Historical log contains detailed information about the data, network characteristics, application level parameters, mobile device information, and external traffic status. 
Data information contains file size ($fs$) and number of files ($n_{files}$). 
Network characteristics contain round trip time ($t_{rtt}$), tcp buffer size ($bs_{tcp}$), and bandwidth ($bw$). 
Application level parameters contain concurrency ($cc$), parallelism ($p$) and I/O block size ($b_{io}$). 
Resource usage information contains CPU utilization ($\mu_{cpu}$), memory utilization ($\mu_{mem}$), NIC card utilization ($\mu_{nic}$), and power consumption ($pw$). 
Mobile device information contains the model, operating system, processor, memory, and network interface (WiFi/radio) specifics. 

Historical logs might contain information about the transfers which were aborted or failed; or sometimes, due to a system error, logs might contain unreasonable information such as achieved throughput greater than the bandwidth. During preprocessing phase we remove those logs. 
{\em Standard outlier detection} model is used to remove those outliers. 

\vspace{1mm}
\noindent {\bf Step 2 -- Clustering similar logs:}
Application level parameters have different impact on different types of transfers. Categorize logs into groups based on their similarity could provide us more structured view of the log information. After analyzing the logs we come to the conclusion that some parameters have direct precedence over other parameters. We use {\em Hierarchical Agglomerative Clustering} which is the most suitable clustering technique for such cases. 


\vspace{1mm}
\noindent {\bf Step 3 -- Optimization:}
This is the most important part of the analysis. We first modeled both throughput and energy function based on historical log. Then we performed mathematical optimization to find the best parameter settings. The details of the optimization are presented in Section~\ref{sec:optimization}.

\vspace{1mm}
\noindent {\bf Step 4 -- Learning optimal parameters:}
As we decided to do historical log analysis on the edge servers or in the cloud, there should be an efficient way to transfer the acquired knowledge from the analysis server to the mobile devices. The simple solution is sending the optimal results gained for each transfer to the mobile device. But, this is not a scalable solution as this approach is too specific to the individual transfers. A mobile device cannot generalize the knowledge for even similar transfers. Moreover, it will take a considerable amount of memory to store those individual results. Another solution would be the use of machine learning techniques, which can be used to learn the knowledge from the optimization step, and can have the power to predict parameters for the unknown transfers. 

Machine learning techniques come with two distinct steps~-- (1) learning and (2) prediction. These two steps can be decoupled. As all the historical logs and optimization results are stored in HLA servers, it is reasonable to do learning step in the HLA server. Then the trained model is transferred to the mobile device. Another reason to choose the number of parameters (also known as weights, connections) in the learning module is fixed. The number of connections and weights do not increase as the historical log increases, only the values of the weights are updated as the model learns. Therefore, HLA server always sends a fixed sized update (e.g., updated values of the weights) to the mobile device. It simultaneously optimizes the memory and communication overhead between the server and the device. In our model we have used off the shelf non-linear machine learning techniques, such as - Artificial Neural Networks (ANN) and Support Vector Machines (SVM). 
As we have limited feature space (number of meta-data in the log), we do not need any deep learning techniques capable of extracting complex pattern from high number of features.

\vspace{1mm}
\noindent {\bf Step 5 -- Scheduling mixed sized data:}
We observed that the files with different sizes can have different optimal parameter settings. Therefore, a dataset containing different sized files should not be transferred with the same parameter settings. A more fine-tuned solution is to cluster the files based on similarity and use optimal parameter settings for each cluster. However, each optimal parameter setting is optimized for that specific cluster and agnostic towards other clusters' parameters. Transferring these clusters concurrently can over-provision the network and introduce packet loss. Therefore, we scale down the parameter values according the the cluster size and some known heuristics. An overview is provided in Algorithm~\ref{algo:scheduling}.

\IncMargin{1em}
\begin{algorithm}[h]
\small
	\SetKw{in}{in}
	\SetKwData{Left}{left}
    \SetKwData{This}{this}
    \SetKwData{Up}{up}
    \SetKwData{Log}{Log}
    \SetKwData{UserLimit}{user\_limit}
    \SetKwData{ismedian}{$e_{s,median}$}
    \SetKwFunction{Median}{Median}
    \SetKwFunction{Union}{Union}
    \SetKwFunction{FindCompress}{FindCompress}
    \SetKwFunction{FindClosestSurface}{FindClosestSurface}
    \SetKwFunction{GetOptimalParam}{GetOptimalParam}
    \SetKwFunction{DataTransfer}{DataTransfer}
    \SetKwFunction{Scheduling}{Scheduling}
    \SetKwFunction{append}{append}
    \SetKwFunction{remove}{remove}
    \SetKwFunction{QueryDB}{QueryDB}
    \SetKwFunction{Sort}{Sort}
    \SetKwFunction{GetSamples}{GetSamples}
    \SetKwFunction{cluster}{cluster}
    \SetKwFunction{GetOptimalParams}{get\_optimal\_params}
    \SetKwFunction{add}{sum}
	\SetKwInOut{Input}{input}\SetKwInOut{Output}{output}
	\Input{Data arguments, $data\_args = \{Dataset , avg\_file\_size, num\_files\}$}
	\Output{Optimal parameter settings, $\theta_{opt}$}
	\BlankLine
    \SetKwProg{proc}{procedure}{}{}
    \proc{\Scheduling{$F_s$, $R_s$, $I_s$} }{
    	$C$ $\leftarrow$ \cluster{$Dataset$}  \\
        \For{$c_{i}$ \in C}{
        	$\theta_i$ $\leftarrow$ \GetOptimalParams($c_i$)     
        } 
        
        \If {\add($\theta_i$) $>$ \UserLimit}{
        	$\theta_{opt}$ $\leftarrow$ $(\theta_i \times \UserLimit) / \add(\theta_i)$ 
        }
    }
\caption{Mixed Data Scheduling}
\label{algo:scheduling}
\end{algorithm}\DecMargin{1em}

\subsection{Optimization}
\label{sec:optimization}

Application level parameters, such as concurrency ($cc$), parallelism ($p$) and I/O block size ($bs$) can be tuned properly to achieve both high throughput and low energy consumptions. 
We define the throughput ($th$) and energy consumption ($\mathbb{E}$) as:
\begin{equation}
th = f_{th}(p,cc,bs)
\end{equation}
\begin{equation}
\mathbb{E} = f_{e}(p,cc)
\end{equation}

Appropriate modeling of $th$ and $\mathbb{E}$ is crucial to find the optimal parameter settings. Historical log contain samples of the parameter space, therefore, can not provide overall view of the whole parameter space. We need an interpolation technique to predict the missing parameters. Then we can optimize these functions to get optimal parameters. We follow two steps - (1) interpolation of unknown parameters, and (2) finding the optimal parameters. These steps are explained below.

\vspace{1mm}
\noindent {\bf Step 1 -- Interpolation of unknown parameters:  }
We observed that the throughput and energy consumption follow a cubic pattern. Therefore, we modeled both throughput and energy consumption as piece-wise cubic interpolation. Cubic interpolation fills the achievable throughput and energy consumption of the unknown parameters. These piece-wise cubic functions are stitched with a guarantee of smoothness up to second derivate. As I/O block size is different from concurrency and parallelism, we modeled it separately. 

To model the throughput, we construct a $2$-dimension cubic spline interpolation for $th = f(bs)$. Piece-wise cubic interpolation can be constructed using interpolant
$th_i = f(bs_i)$ and connecting them by maintaining smoothness up to second derivative. we can define each cubic polynomial piece as generic cubic function: 

\begin{equation}
f_i(bs) = x_{i,0} + x_{i,1}bs + x_{i,2}bs^2 + x_{i,3}bs^3,\forall bs \in [bs_i,bs_{i+1}].
\end{equation}

Boundaries can be constrained as $f(bs_{i+1}) = f(bs_i)$. We can have:
\begin{equation}
f_i(bs_i) = th_i, \quad i = 1,...,N
\end{equation}
Therefore, the $N$ continuity constraints of $f(bs)$ are as:
\begin{equation} \label{eq:constraint2}
f_{i-1}(bs_i) = th_i = f_{i}(bs_i), \quad i = 2,...,N.
\end{equation}

The following constraint confirms smoothness up to second derivatives. 
\begin{equation} \label{eq:constraint3}
\dfrac{d^2 f_{i-1}}{d^2 bs} (bs_i) = \dfrac{d^2 f_{i}}{d^2 bs} (bs_i), \quad i = 2,...,N
\end{equation}
The boundary condition for spline could be written as:
\begin{equation} \label{eq:constraint4}
\dfrac{d^2 f}{d^2 bs} (bs_1) =  \dfrac{d^2 f}{d^2 bs} (bs_n) = 0
\end{equation}
The coefficients can be computed by solving the system of linear equations. 

Throughput is also dependent on concurrency and parallelism. The example above can be extended to generate throughput surface with two independent variables - $cc$ and $p$. Similarly, we modeled energy as a function of concurrency, parallelism and I/O block size. 

\vspace{1mm}
\noindent {\bf Step 2 -- Find optimal parameters: }
Energy efficient transfer aims to reduce the energy consumption without compromising the transfer performance. This objective function tries to optimize both throughput and power consumption at the same time. This objective function does not guarantee both maximally achievable throughput with minimum power consumption, however, it ensures that every unit of power can be spent to achieve highest possible throughput under the energy efficiency constraint. The objective function here is to maximize achievable throughput over power consumption. 
\begin{equation}
\mathrm{maximize} \quad \int_{t_s}^{t_f} th/ \mathbb{E}
\end{equation}

We take into account all the boundary constraints for the parameters and other necessary constraints. Due to the space limitation we are not including those here. Then we used non-linear optimizer to find the optimal parameters. 

\subsection{Evaluation of FastHLA Model}
\label{sec:experiment}

\begin{figure*}[t]
    \begin{centering}
    \begin{subfigure}[t]{0.32\textwidth}
        \includegraphics[keepaspectratio=true,width=58mm]{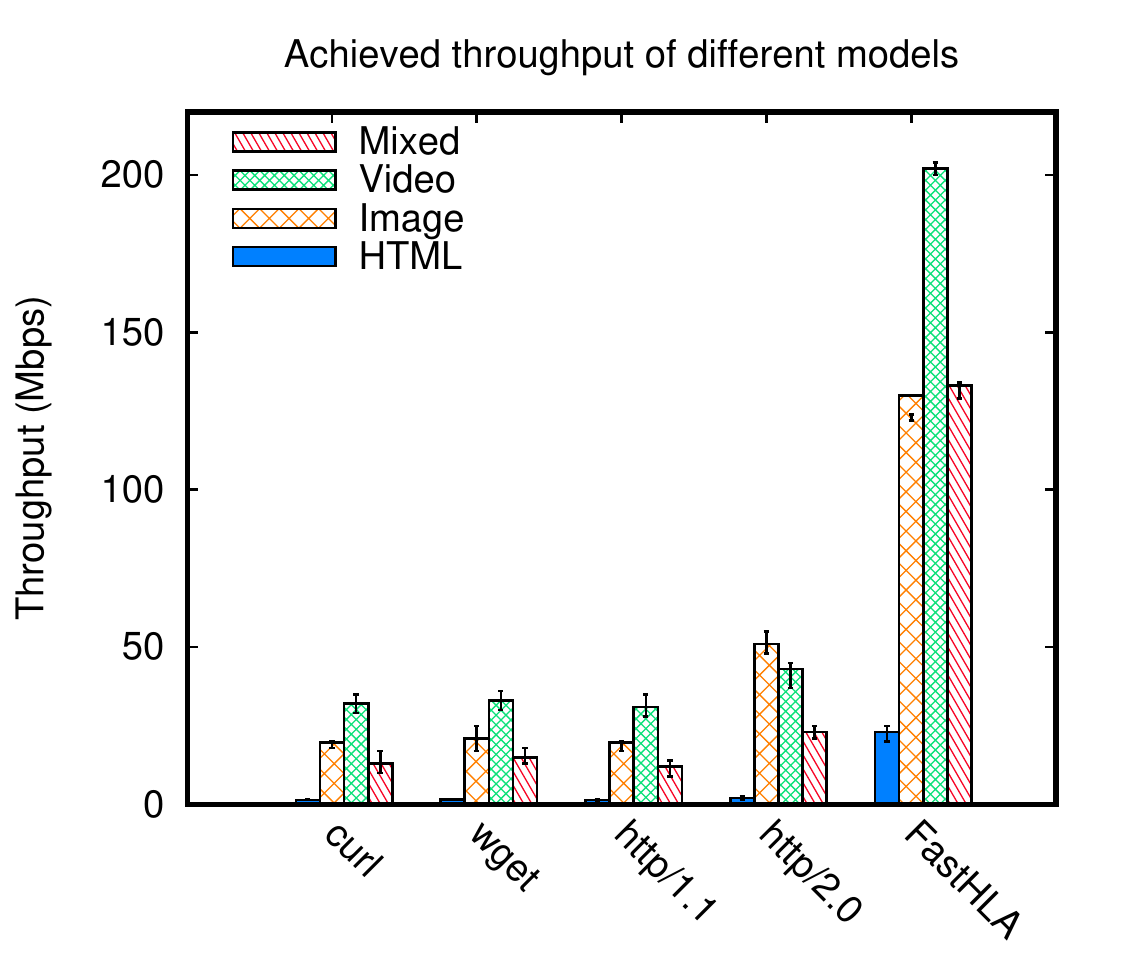}
        \vspace{-5mm}
        \caption{Achieved Throughput}
    \end{subfigure}
    ~
    \begin{subfigure}[t]{0.32\textwidth}
        \includegraphics[keepaspectratio=true,width=58mm]{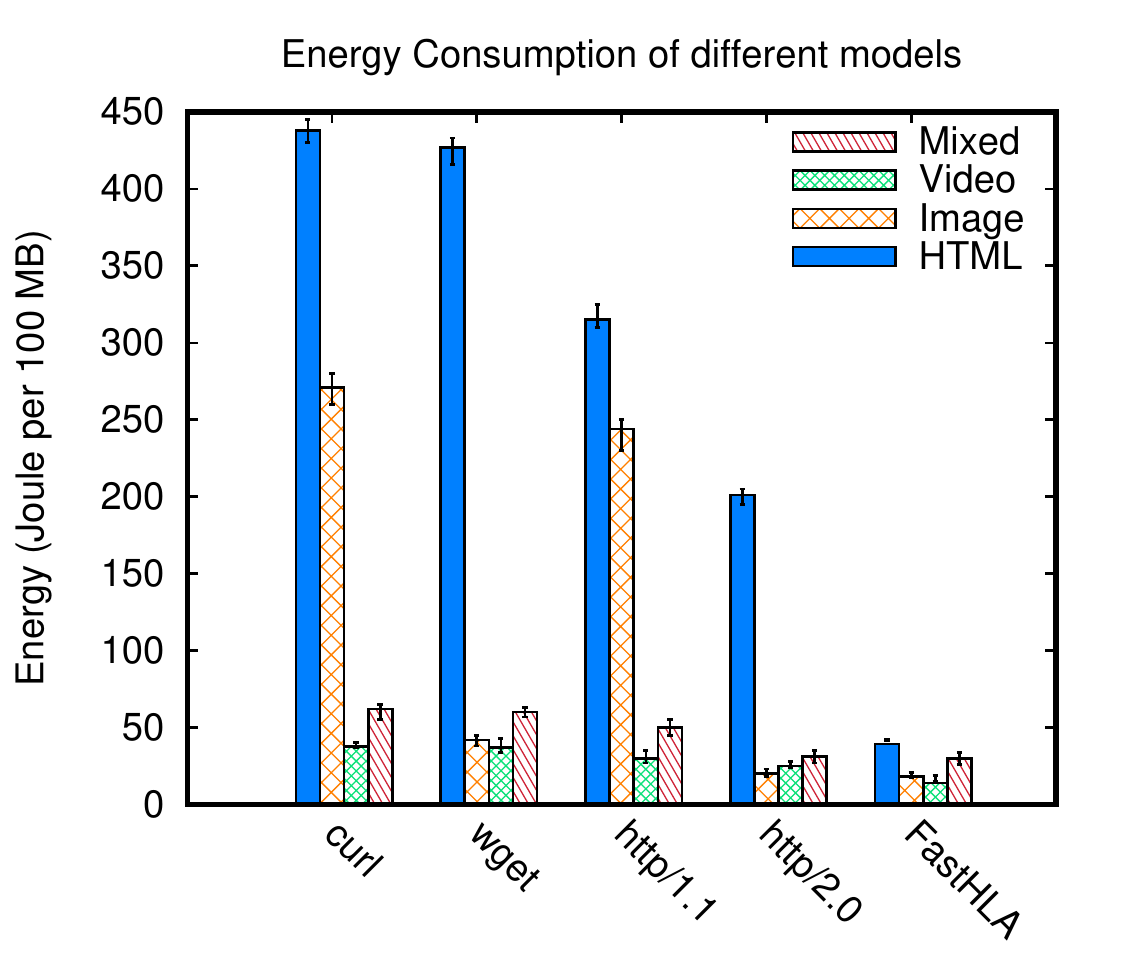}
        \vspace{-5mm}
        \caption{Energy Consumption}
    \end{subfigure}
    ~
    \begin{subfigure}[t]{0.32\textwidth}
      \includegraphics[keepaspectratio=true,width=58mm]{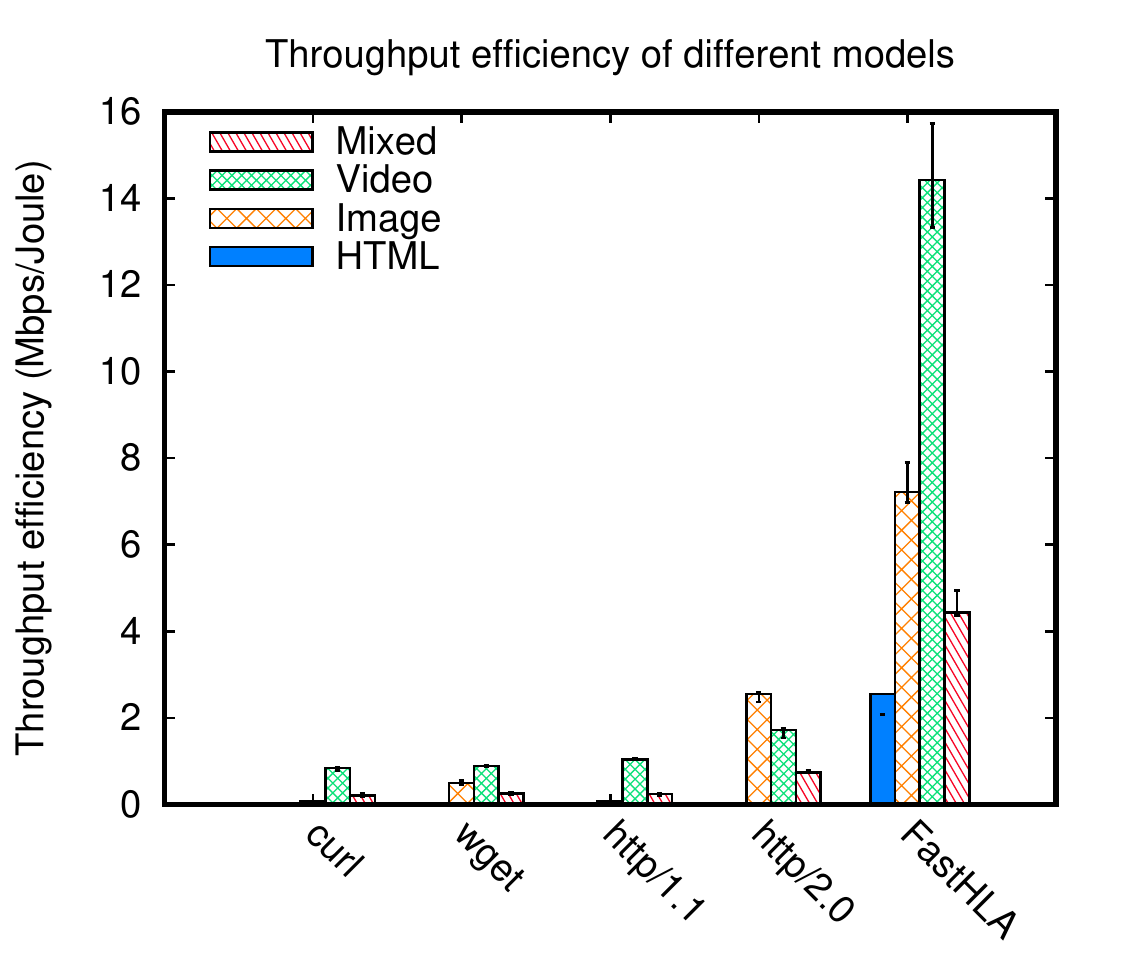}
       \vspace{-5mm}
        \caption{Throughput Efficiency}
    \end{subfigure}
    \vspace{-8mm}
     \caption{Achievable throughput and corresponding energy consumption of different optimization objectives and the accuracy of the learning module.}
     \vspace{-4mm}
     \label{fig:objective_comparison}
     \end{centering}
 \end{figure*}
 
\begin{figure}[t]
\begin{centering}
\includegraphics[keepaspectratio=true,angle=0,width=58mm]{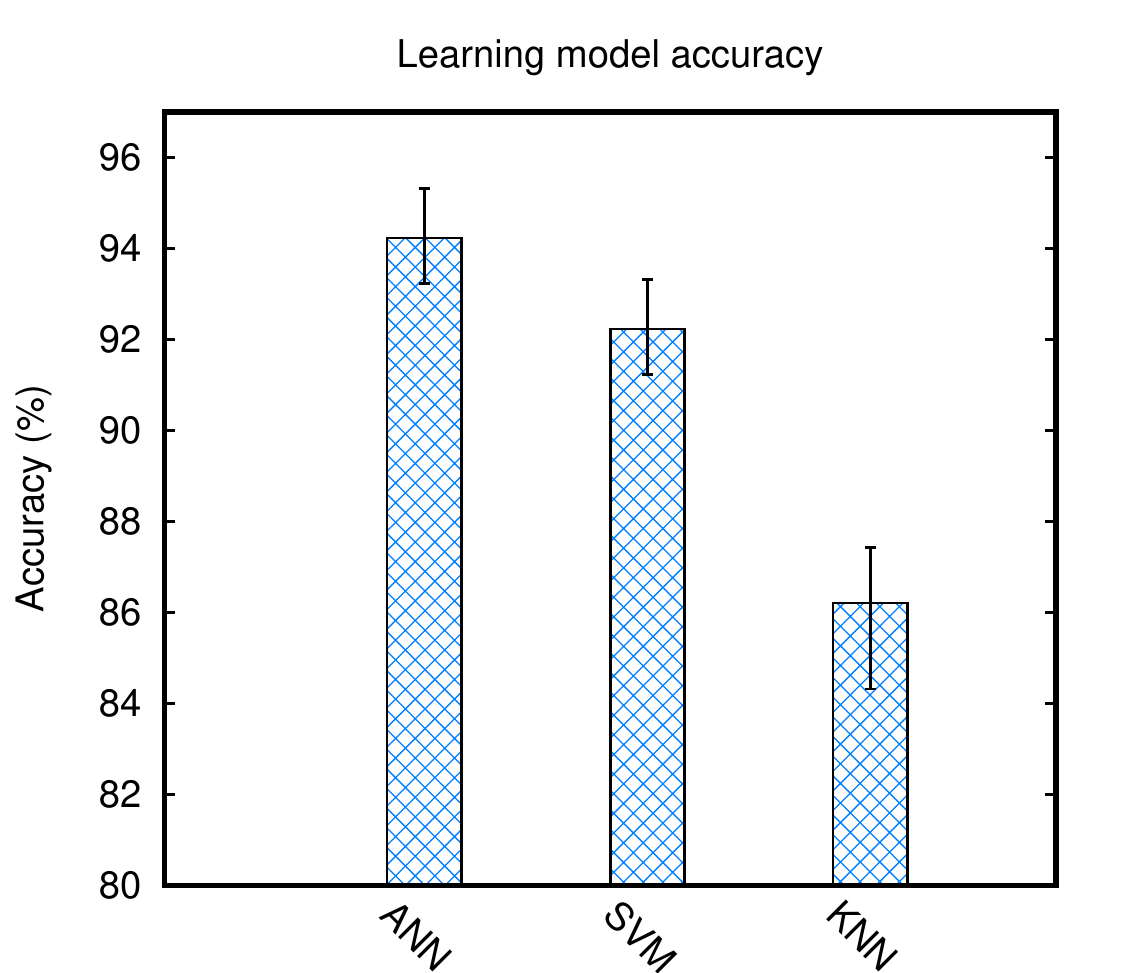}
	\end{centering}
    \vspace{-3mm}
\caption{Accuracy of different learning modules.} 
\label{fig:LM_accuracy}
\vspace{-3mm}
\end{figure}

Historical log analysis (HLA) data transfer experiments are conducted in the same experimental testbed described in Section~\ref{sec:methodology}.
We trained and tested our model (FastHLA) on the real data transfer logs and compared the performance and power consumption FastHLA with energy-agnostic wget~\cite{wget} and curl~\cite{curl} clients as well as two versions of de-facto application layer transfer protocol of HTTP, which are HTTP/1.1 and HTTP/2~\cite{http2}. While HTTP/1.1 is a textual protocol, the newly introduced HTTP/2 is a binary protocol that supports multiplexing, header compressions and lets the server to {\em push} responses. 

To evaluate our model, we used HTML, image and video datasets (as described in Section~\ref{sec:methodology}) along with a combined dataset that contains a mix of three datasets. We compared FastHLA with other models using these dataset so that we can get a fine-grained analysis of performance. Figure~\ref{fig:objective_comparison} shows both energy and throughput comparison of different existing approaches along with our model. We observe that FastHLA outperforms all other tested solutions in every data category. For image, video, HTML, and mixed data sets, we see  2$\times$, 4$\times$, 10$\times$ and 4$\times$ throughput improvement over the closest competitor HTTP/2. HTTP/2 uses multiplexing to transfer multiple streams over a single connection to remove head-of-line blocking. However, single connection can achieve very poor results in long RTT WAN links. HTTP/1.1 uses multiple connections to request multiple files, however, there is no way to dynamically set those number of connections (parallelism). On the other hand, we have used historical analysis to decide on optimal level of parallelism, concurrency and I/O block size. 

Figure~\ref{fig:objective_comparison}(b) shows the energy consumption of different models. As we can see standard applications like wget or curl are not optimized for power consumption and draw a huge energy compare to HTTP/2. On the other hand, FastHLA improves power consumption 5$\times$ and 2$\times$ for HTML and video files respectively compared to HTTP/2. We have observed that high transfer throughput can shorten the data transfer time. That means CPU has to work for a shorter period of time and CPU consumes most of the power during the transfer. That explains why FastHLA consumes less power compared to other approaches.  However, the energy consumption is similar for image and mixed data. Even if the power consumption is similar for image and video, FastHLA can provide more achievable throughput compared to HTTP/2. We use throughput efficiency, $th/\mathbb{E}$ to measure the energy efficiency of the models (as shown in Figure~\ref{fig:objective_comparison}(c)). FastHLA improves throughput efficiency  2.5$\times$ for both video and image data.  

We have used three different learning modules to see the efficiency of those models. Among them ANN and SVM can reach up to 94\% and 92\% accuracy respectively (as shown in Figure~\ref{fig:LM_accuracy}). However, KNN can achieve up to 86\% accuracy. As K-Nearest Neighbor takes into account $k$ closest logs to decide on the parameters, it is not feasible to transfer all the optimal results to mobile due to memory issues. That is why we decided not to use KNN in our model. However, the shallow Neural Network and SVM both can learn efficiently the optimal parameters. We can use coefficient of determination, $R^{2}$ to see how well the the prediction learning module matches the target. It can be expressed as:

\begin{equation}
R^{2} = 1 - \sum(y_{i} - \bar{y})^{2} / \sum(f_{i} - \bar{y})^{2}
\end{equation}

where $y_{i}$ is the actual optimal throughput and $\bar{y}$ is the mean of $y_{i}$. The predicted throughput is defined as $f_{i}$. $R^{2}$ is a good statistical indicator that can point out the strongly actual and prediction values are related. In case of perfect matches between all known targets and the predictions $R^{2}$ value will be 1. However, we can say a model can predict with good generalization if $R^{2}$ value is close to 1. We computed this for both ANN and SVM as $R^{2}_{ANN} = 0.92$ and  $R^{2}_{SVM} = 0.87$ respectively. 

\section{Conclusion \& Future Work}
\label{sec:conclusion}

In this paper, we performed extensive analysis and presented the effects of application-layer data transfer protocol parameters (such as the number of parallel data streams per file, the level of concurrent file transfers to fill the mobile network pipes, and the I/O request size) on mobile data transfer throughput and energy consumption for WiFi and 4G LTE connections. 
We also proposed a novel historical-data analysis based model, called FastHLA, that can achieve significant energy savings at the application layer during mobile network I/O without sacrificing the performance. 
Our analysis shows that FastHLA model can achieve significant energy savings using only application-layer solutions at the mobile systems during data transfer with no performance penalty. We also show that, in many cases, our FastHLA model can increase the performance and save energy simultaneously.

According to our experiments, by intelligently tuning the concurrency and parallelism levels during data transfers, our FastHLA model can increase the data transfer throughput by up to 10X, and decrease the energy consumption by up to 5X compared to state-of-the-art HTTP/2.0 transfers. The improvement is even larger compared to base HTTP/1.1 transfers and client tools such as wget and curl. 

As a future work, we are planning to develop SLA-based transfer tuning algorithms to balance the performance vs energy trade-off during mobile network I/O according to the preferences of the mobile users. 
While keeping the quality of service (i.e., transfer throughput) at the desired level, these algorithms will try to keep the energy consumption at the minimum possible level. 

\bibliographystyle{abbrv}
\bibliography{main}

\end{document}